%% file: email.tex
\documentclass[conference]{IEEEtran}
\pdfoutput=1

\usepackage{booktabs} 
\usepackage{balance}
\usepackage{multirow}
\usepackage{epsfig}
\usepackage{subfigure}
\usepackage{hyperref}
\usepackage{amsmath, amssymb, amsfonts}
\usepackage{color}
\usepackage{subfigure}
\usepackage{textcomp}
\usepackage{listings}
\usepackage{url}
\usepackage{pifont}
\usepackage{multicol}
\usepackage{graphicx}
\usepackage{tabu}
\usepackage{tikz}
\usepackage{xcolor}
\usepackage{colortbl}
\usepackage[export]{adjustbox}
\usepackage{array}
\newcolumntype{P}[1]{>{\centering\arraybackslash}p{#1}}

\DeclareRobustCommand{\cmark}{$\checkmark$}%
\DeclareRobustCommand{\xmark}{\ding{53}}%

\DeclareRobustCommand{\block}{\tikz[baseline=-0.5ex]\draw[black,fill=black,radius=2pt] (0,0) circle;}%
\DeclareRobustCommand{\inbox}{\tikz[baseline=-0.5ex]\draw[black,fill=white,radius=2pt] (0,0) circle;}%
\DeclareRobustCommand{\spam}{\tikz[baseline=-0.5ex]\draw[black,fill=black,even odd rule] (0,0) circle (2pt) (0,0) circle (1pt);}%

\newenvironment{packed_itemize}{
\begin{list}{\labelitemi}{\leftmargin=1.5em}
\setlength{\itemsep}{4pt}
\setlength{\parskip}{0pt}
\setlength{\parsep}{0pt}
\setlength{\headsep}{0pt}
\setlength{\topskip}{0pt}
\setlength{\topmargin}{0pt}
\setlength{\topsep}{0pt}
\setlength{\partopsep}{0pt}
}{\end{list}}

\newcommand{\para}[1]{{\vspace{4pt} \bf \noindent #1
\hspace{10pt}}}

\newcommand{\eg}{{\it e.g.}}

\begin{document}



\title{Revisiting Email Spoofing Attacks}

\author{
{\rm Hang Hu}\\
Virginia Tech
\and
{\rm Gang Wang}\\
Virginia Tech
}



\maketitle

\begin{abstract}
The email system is the central battleground against phishing and
social engineering attacks, and yet email providers still face key
challenges to authenticate incoming emails. As a result,
attackers can apply spoofing techniques to impersonate a trusted
entity to conduct highly deceptive phishing attacks. In this work, we study
{\em email spoofing} to answer three key questions: (1)
How do email providers detect and handle forged emails? (2)
Under what conditions can forged emails penetrate the defense to reach
user inbox? (3) Once the forged email gets in, how email providers
warn users? Is the warning truly effective?

We answer these questions through end-to-end measurements on 35 popular email
providers (used by billions of users), and extensive user studies ($N=913$)
that consist of both simulated and real-world phishing experiments. We
have four key findings. {\em First}, most
popular email providers have the necessary protocols to detect spoofing, but still
allow forged emails to get into user inbox ({\em e.g.}, Yahoo Mail,
iCloud, Gmail). {\em Second}, once a forged email gets in, most email
providers have no warnings for users, particularly on mobile email
apps. Some providers ({\em e.g.}, Gmail Inbox) even have
misleading UIs that make the forged email look authentic. {\em
  Third}, a few email providers (9/35) have implemented visual
security cues for unverified emails, which demonstrate a positive
impact to reduce risky user actions. Comparing simulated experiments
with realistic phishing tests, we
observe that the impact of security cue is less significant when users
are caught off guard in the real-world setting.

\end{abstract}

\input{introduction}

\input{background}
\input{result}
\input{user}

\input{discuss}

\input{related}
\input{conclusions}
\begin{small}
\bibliographystyle{IEEEtranS}
\bibliography{bibliography.bib,astro.bib,wang.bib}
\end{small}

\input{appendix}

\end{document}

%% file: introduction.tex
\section{Introduction}
\label{sec:intro}
Despite the recent development of system and network security, human
factors still remain the weak link. As a result, attackers
increasingly rely on phishing and social
engineering tactics to breach their targeted networks. In 2016
alone, email phishing has involved in nearly half of the 2000+
reported security breaches, causing a leakage of billions of user
records~\cite{verizon17, ccs17_breaches}.

{\em Email spoofing} is a critical step in phishing, where the attacker
impersonates a trusted entity 
to gain the victim's trust. 
Unfortunately, today's email transmission protocol (SMTP) has no built-in mechanism to
prevent spoofing~\cite{smtp}. It relies on email providers to
implement SMTP extensions such as SPF~\cite{spf}, DKIM~\cite{dkim} and
DMARC~\cite{dmarc} to authenticate the sender. Since implementing these extensions is {\em voluntary},
their adoption rate is far from satisfying. Recent measurements
conducted in 2015 show that among Alexa top 1
million domains, 40\% have SPF, 1\% have DMARC, and even
fewer are correctly/strictly configured~\cite{Durumeric:2015,Foster:2015}.





The limited server-side protection is likely to put {\em users} in a
vulnerable position. Because not every sender
domain has adopted SPF/DKIM/DMARC, email providers still face
key challenges to reliably authenticate incoming emails.
When an email failed the authentication, it is a
``blackbox'' process in terms of how email providers handle this email.
 Would forged emails still be delivered to users? If so, how could
 users know the email is questionable? Take Gmail for example, Gmail delivers
certain forged emails to the inbox and places a security cue on the
sender icon (a red question mark, Figure~\ref{fig:screen}(a)). We are
curious about how a broader range
of email providers handle forged emails, and how
much the security cues actually help to protect users.

In this paper, we describe our efforts and experience in evaluating the effectiveness
of user-level protections against email spoofing\footnote{Our study has been reviewed and approved by our local IRB
  (IRB-17-397).}. We answer the above questions in two key
steps. {\em First}, we examine how popular email providers detect and handle
forged emails using end-to-end measurements. Particularly, we examine under what
conditions can forged/phishing emails reach the user inbox and what
security cues (if any) are used to warn users. {\em Second}, we
measure the impact of the security cues on users through a series of
user studies ($N=913$). To obtain reliable experiment results, we conduct both simulated
and real-world phishing experiments.




\para{Measurements.} By scanning Alexa top 1 million hosts in January
and October 2017, we confirm that the adoption rate of SMTP security
extensions is still low (SPF 45.0\%, DMARC 4.6\%).
This motivates us to examine how email providers handle incoming emails that
failed the authentication. We conduct end-to-end spoofing experiments
on 35 popular public email providers used by billions of
users. We treat each email provider as a blackbox and vary the input
(forged emails) to monitor the output (the receiver's inbox).

We find that forged emails can penetrate the majority of email
providers (33/35) including Gmail,
Yahoo Mail and Apple iCloud under proper conditions. Even if the
receiver performs all the authentication checks (SPF, DKIM,
DMARC), spoofing an unprotected domain or a domain with ``weak''
DMARC policies can help the forged email to reach the inbox. In addition, spoofing
an ``existing contact'' of the victim also helps the
attacker to penetrate email providers ({\em e.g.}, Hotmail). Overall, the result indicates that providers tend to prioritize
email delivery when the email cannot be reliably authenticated.


More surprisingly, while most providers allow forged emails to
get in, rarely do they warn users of the unverified sender.
Only 9 of 35 providers have implemented some security cues: 8
providers have security cues on their
{\em web interface} ({\em e.g.}, Gmail) and only 4 providers ({\em e.g.}, Naver) have the security cues consistently for the {\em mobile apps}.
Even worse, certain email providers have misleading UI elements which
help the attacker to make forged emails look authentic.
For example, when attackers spoof an existing contact (or a user from the same provider),
25 out of 35 providers will automatically load either the sender's
profile photo, name card or email history from their internal database
for this forged email. These designs are supposed to improve email
usability by providing the context on ``who sent this
email''. However, when the sender address is actually spoofed, these designs, in turn,
help the attacker with the deception.

\para{User Studies.} While a handful of email providers have
implemented security cues, the real question is how
effective they are. 
We answer this question using a series of user studies ($N=913$) where
participants examine spoofed phishing emails with or without security
cues on the interface. Our methodology follows a hybrid
approach. First, we conducted a ``role-playing'' where users examine phishing and benign emails in a
controlled hypothetical scenario. While the role-playing method is widely used
for usability studies~\cite{jJPMB12, Sheng:2010:, Vishwanath:2011, Wang:2016,
  Wu:2006:}, our concern is that certain artificial conditions may
affect the user behavior. To this end, we also run a real-world
phishing test to validate our results. 


Our result shows that security cues have a positive impact on reducing
risky user actions consistently across the role-playing experiment and
the real-world phishing test. The impact is also consistently positive
for users of different demographics (age, gender, education-level). 
In addition, comparing with the role-playing setting, we find that the
impact of security cues in the real-world
setting is not as significant when users are caught off guard. The
result indicates that the security cue is useful but cannot eliminate
the phishing risk. The design of security cue also should be further improved to
maximize its impact. 

\para{Contributions.} Our work has three key contributions:
\begin{itemize}
\item {\em First}, our end-to-end measurement sheds light on how email
  providers handle forged emails. We reveal the
  trade-offs made by different email providers between email availability and security.

\item {\em Second}, we are the first to empirically analyze the usage of security cues on
  spoofed emails. We show that most email providers not only lack the
  necessary security cues (particularly on mobile apps), but introduce
  misleading UIs that in turn help the attackers.

\item {\em Third}, we conduct extensive user studies to evaluate the
  effectiveness of the security cue. We demonstrate its positive impact
  (and potential problems) using both role-playing experiments and real-world
  phishing tests. 
\end{itemize}

%% file: background.tex
\section{Background and Methodology}
\label{sec:back}
Today's email system is built upon the SMTP protocol, which was
initially designed without security in mind. {\em Security extensions}
were introduced later to provide confidentiality, integrity, and
authenticity. In this paper, we primarily focus on
email authenticity, a property that is often abused by phishing
attacks. Below, we briefly introduce SMTP and related security
extensions. Then we discuss the key
challenges to defend against email spoofing and introduce our
research questions and methodology.

\begin{figure}
         \centering
      	\begin{minipage}[t]{0.45\textwidth}
   	\includegraphics[width=0.99\textwidth]{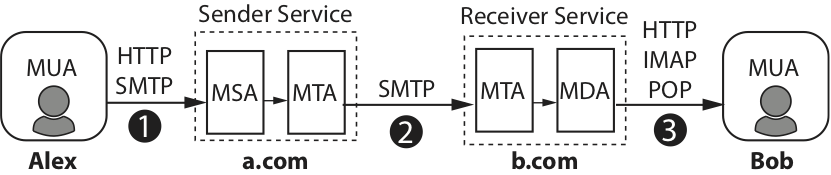}
	\end{minipage}
\caption{Email transmission from Alex to Bob.}
	\label{fig:sys}
  	\vspace{-0.18in}
\end{figure}

\subsection{SMTP and Email Spoofing}
Simple Mail Transfer Protocol (SMTP) is an Internet standard for
electronic mail transmission~\cite{smtp}. Figure~\ref{fig:sys} shows
the three main steps to deliver an email message. (\ding{182})
Starting from the sender's Mail User Agent (MUA), the message is first transmitted to the Mail
Submission Agent (MSA) of the sender's service provider via STMP
or even HTTP/HTTPS. (\ding{183}) Then the sender's Mail Transfer Agent (MTA)
sends the message to the receiver's email
provider using SMTP. (\ding{184}) After processing, the message is delivered to the
receiving user by the Mail Delivery Agent (MDA)
via Internet Message Access Protocol (IMAP), Post Office
Protocol (POP) or even HTTP/HTTPS.

When initially designed, SMTP did not have
any security mechanisms to authenticate the
sender identity. As a result, attackers can easily craft a forged email to
impersonate/spoof an arbitrary sender address by modifying the ``{\tt MAIL
  FROM}'' field in SMTP. Email spoofing is a critical step in a
phishing attack --- by impersonating a trusted entity as the email
sender, the attacker has a higher chance to gain the victim's
trust. In practice, attackers usually exploit SMTP in step (\ding{183}) by setting up
their own MTA server.


\subsection{Email Authentication}
To provide security for the email system, various security
extensions have been proposed and standardized. Here, we briefly
describe the extensions that are designed to support email
authenticity and prevent spoofing (\eg, SPF, DKIM and DMARC). There
are other security extensions
that provide confidentiality and integrity (\eg,
STARTTLS), which are beyond the scope of this paper.

\para{SPF.}
Sender Policy Framework (SPF) allows an email service (or an
organization) to publish a list of IPs that are authorized to
send emails for its domain (RFC7208~\cite{spf}). For example, if
a domain ``{\tt a.com}'' published its SPF record in the DNS, then other
email services can check this record when receiving an email that claims to
be from ``{\tt \{someone\}@a.com}''. In this way, only authorized IPs
can send emails as ``{\tt a.com}''. SPF makes it convenient for an
organization to use third-party/cloud-based email
providers by simply adding the provider's IP range to the SPF
record.  In addition, SPF allows the organization to specify a policy regarding how the
receiver should handle the email that failed the authentication.


\para{DKIM.}
DomainKeys Identified Mail (DKIM) uses the public-key based approach to
authenticate the email sender (RFC6376~\cite{dkim}). The sender's
email service will place a digital signature in the email
header signed by the private key associated
to the sender's domain. The receiving service can retrieve the
sender's public key from DNS to verify the signature.
In order to query a DKIM public key from DNS, one not only
needs the domain name but also a {\em
  selector} (an attribute in the DKIM
signature). Selectors are used to permit multiple keys under the same
domain for more a fine-grained signatory control. DKIM does not specify what actions that the receiver should
take if the authentication fails.


\para{DMARC.}
Domain-based Message Authentication, Reporting and Conformance (DMARC)
is built on top of SPF and DKIM (RFC7489~\cite{dmarc}).
DMARC is not a standalone protocol. It allows the domain
administrative owner to publish a policy to specify 
 what actions the receiver should take when the incoming email fails the SPF and DKIM check. A domain's DMARC
record is available under {\tt \_dmarc.domain.com} in DNS. Another
protocol called ADSP (Author Domain Signing Practices) also allows
domain owners to publish DKIM signing policies, but ADSP was declared
``historic'' in 2013~\cite{adsp}.



\subsection{Practical Challenges to Prevent Spoofing}
In theory, we already have the tools to secure the email
system against spoofing. However, significant
challenges remain when these tools are not properly
deployed in practice. Recent measurements conducted in 2015 show that the
adoption rates of SMTP security extensions are far from
satisfying~\cite{Durumeric:2015, Foster:2015}. Among Alexa top 1 million
domains, only 40\% have published an SPF
record, and an extremely low percentage of domains (1\%) have a DMARC
policy. In addition, the majority of policies are not strictly reject
(``reject'' policy means the receiver should reject the email if it
failed the authentication~\cite{Foster:2015}).

These results indicate a real challenge to protect users
against email spoofing. First, with a large number of domains
not publishing an SPF/DKIM record, email providers cannot reliably detect
incoming emails that spoof unprotected domains. Second, even a
domain is SPF/DKIM-protected, the lack of (strict) DMARC policies puts
the receiving server in a difficult position. Without a clear instruction to
reject these emails, what would the receiving email providers do? 
Existing works mainly examined the authentication protocols on the
{\em server-side}. However, there is still a big gap
between the server-side spoofing
detection and the actual impact to users.

\subsection{Research Questions and Method}
Our study seeks to understand email spoofing and related protections from
the {\em user perspectives}. We have three key questions.
(1) When email providers face uncertainty in authenticating
incoming emails, how would they handle the situation? Under what
conditions would forged emails be delivered to the inbox? (2) Once forged
emails get in, what types of warning mechanisms (if any) are used to
notify users of the unverified sender address? (3) How
effective are existing warning mechanisms to protect users? Answering these questions
is critical to understanding the {\em actual risks} that are exposed to
users by email spoofing.


\para{Methods.} We answer question (1)--(2) through end-to-end
spoofing experiments (\S\ref{sec:1m}, \S\ref{sec:spoof} and
\S\ref{sec:result}). For a given email provider, we treat it as a ``blackbox''. By controlling the
input ({\em e.g.}, forged emails) and
monitoring the output (receiver's inbox), we
infer the decision-making process inside the blackbox. 
In addition, we empirically examine
the usage of visual security cues to warn users of forged emails.


We answer question (3) by conducting a large user
study (\S\ref{sec:design} and \S\ref{sec:user}). The idea is to let
users read spoofing/phishing emails with and without security
cues on the interface. Instead of simply applying traditional user
study methods (which has limitations to capture realistic user
behavior~\cite{ieeesp07}), our approach combines controlled
``role-playing'' experiments with ``deceptive'' real-world phishing
tests in order to
produce reliable results.




\para{Ethics.} We have taken active steps to ensure research
ethics. Our measurement study only uses dedicated
email accounts owned by the authors and there is no real user getting
involved. In addition, to minimize the impact on the target email
services, we have carefully controlled the message sending
rate (one message every 10 minutes), which is no
different than a regular email user.
For the user studies, all the experiments have obtained user
consent either before or after the study. Email samples used in the user study are completely
benign (screenshots or emails with benign links). For the user study
that involves ``deception'', we worked closely with IRB to review and revise the
experiment design. 
More detailed ethical discussions will be presented in later sections.

\section{Adoption of SMTP Extensions}
\label{sec:1m}
The high-level goal of our measurement is to provide an end-to-end
view of email spoofing attacks against popular email providers. 
Before doing so, we first examine the recent adoption rate of email security
extensions among Internet domains compared with that from two years
ago~\cite{Durumeric:2015, Foster:2015}. This helps to provide the
context for the challenges that email providers face to
authenticate incoming emails.

\para{Scanning Alexa Top 1 Million Domains.}
Email authentication requires the sender domains to publish their
SPF/DKIM/DMARC records to DNS. 
To examine the recent adoption rate of SPF and DMARC among potential sender
domains, in January and October 2017, we crawled two snapshots of the DNS record for Alexa top 1
million hosts~\cite{Alexa}. Similar to existing work~\cite{Durumeric:2015, Foster:2015},
this measurement cannot apply to DKIM, because querying the DKIM
record requires knowing the {\em selector} information for every each
domain. The selector information is only available in the DKIM
signature in the email header, which is not a public
information. We will measure the DKIM usage later in the
end-to-end experiment.

\begin{table}[t]
\begin{center}
\caption{SPF/DMARC statistics of Alexa 1 million domains. The data was
collected in October 2017. }
\label{tab:1m}
\vspace{-0.08in}
\begin{tabu} to 0.47\textwidth{X[l]X[r]X[r]}
\tabucline[1.1pt]{-}
 Status  & All Domains \# (\%)  & MX Domains \# (\%) \\
\hline
Total domains & 1,000,000 (100\%) &  803,720 (100\%) \\
\hline
w/ SPF & 493,367 (49.3\%) & 475,596 (59.2\%)\\
w/ valid SPF & {\bf 449,848 (45.0\%)} & {\bf 432,669 (53.8\%)}\\
Policy: soft fail & 275,244 (27.5\%) & 270,994 (33.7\%)\\
Policy: hard fail  & {\bf 123,084 (12.3\%)} & {\bf 111,231 (13.8\%)} \\
Policy: neutral & 50,437 {5.0\%} & 49,381 (6.1\%)\\
Policy: pass & 1083(0.1\%) & 1,063 (0.1\%)\\
\hline
w/ DMARC & 46,159 (4.6\%) & 44,318 (5.5\%)\\
w/ valid DMARC & {\bf 45,580 (4.6\%)} & {\bf 43,771 (5.4\%)}\\
Policy: none & 35,363 (3.5\%) &  34,387 (4.3\%)\\
Policy: reject & {\bf 5,614 (0.6\%)} & {\bf 4,875 (0.6\%)}\\
Policy: quarantine & 4,603 (0.5\%) & 4,509 (0.6\%)\\
\tabucline[1.1pt]{-}
\end{tabu}
\end{center}
\vspace{-0.25in}
\end{table}

\para{Results.} Table~\ref{tab:1m} shows the statistics for the
October 2017 snapshot. The results from the January snapshot are very
similar and we refer interested readers to Appendix A. As shown in
Table~\ref{tab:1m}, SPF and DMARC both have some increase in
the adoption rate compared to two years
ago, but the increases are not very significant. About 45.0\% of the domains have published a
valid SPF record in October 2017 (40\% in 2015~\cite{Foster:2015}),
and 4.6\% have a valid DMARC record in 2017 (1.1\% in
2015~\cite{Foster:2015}). The invalid records are often caused by the
domain administrators using the wrong format for the SPF/DMARC
record. Another common error is to have
multiple records for SPF (or DMARC), which is equivalent to ``no record'' according to RFC7489~\cite{dmarc}. 

Among the Alexa top 1 million domains, 803,720 domains are MX
domains ({\em i.e.}, mail exchanger domains that host email
services). The adoption rates among MX domains are slightly higher (SPF 53.8\%,
DMARC 5.4\%). For non-MX domains, we argue that it is also important to adopt the
anti-spoofing protocols. For example, {\tt office.com} is not a MX
domain but it hosts the product web page for the Microsoft Office software. A
potential attacker can spoof {\tt office.com} to phish Microsoft Office users (or even Microsoft
employees). 


SPF and DMARC both specify a policy regarding what actions the
receiver should take after the authentication fails. Table~\ref{tab:1m} shows that only a small portion
of the domains specifies a strict ``reject'' policy: 12.3\% of the
domains set ``hard fail'' for SPF, and 0.6\% set
``reject'' for DMARC. 
The rest of the domains simply leave the decision to the email receiver. 
``Soft fail''/``quarantine'' means that the email
receiver should process the email with caution. ``Neutral''/``none''
means that no policy is specified. SPF's ``pass'' means
that the receiver should let the email go through regardless. 
If a domain has both
SPF and DMARC policies, DMARC overwrites SPF
as long as the DMARC policy is not ``none''.

Domains that use DKIM also need to publish
their policies through DMARC. The fact that only 4.6\% of the domains have a valid DMARC
record and 0.6\% have a ``reject'' policy indicates that most DKIM adopters also did
not specify a strict reject policy.

\section{End-to-End Spoofing Experiments}
\label{sec:spoof}
Given the low adoption rate of SPF and DMARC among Internet domains, it is
still challenging for email providers to reliably authenticate all
incoming emails. When encountering questionable emails, we are curious about how email
providers make the trade-off between email delivery and security.
In the following, we describe our experiment methodology to answer two
key questions: First, under what conditions will email
providers let forged emails get into the user inbox? Second, when it happens, would
users receive any warnings? 

\subsection{Experiment Setup}
\begin{figure}
         \begin{center}
      	\begin{minipage}[t]{0.45\textwidth}
   	\includegraphics[width=0.99\textwidth]{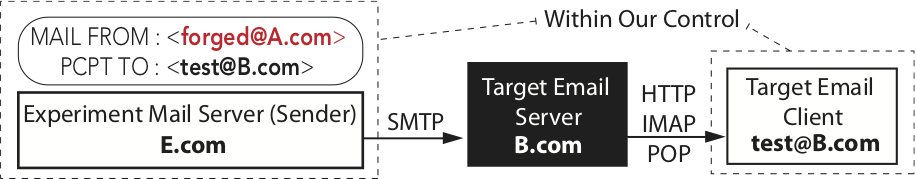}
	\end{minipage}
\caption{End-to-end spoofing experiment. Our server
  is located at {\tt E.com} which sends a forged email to the target
  email service at {\tt B.com}. The email spoofs {\tt A.com} as
  the sender.}
	\label{fig:exp}
  	\vspace{-0.2in}
\end{center}
\end{figure}

We conduct end-to-end spoofing
experiments on popular email providers that are used by billions of
users. The high-level idea is illustrated in Figure~\ref{fig:exp}. For
a given email provider ({\tt B.com}), we put it to the receiving
end. We set up a user account under {\tt B.com} as the email receiver ({\tt
  test@B.com}). Then we set up an experimental
server ({\tt E.com}) to send forged emails to the receiver account.
Our server runs a Postfix mail service~\cite{Postfix} to directly interact with the target mail server using SMTP.
By controlling the input (the forged email) and observing the output
(the receiver account), we infer the decision-making process inside of the
target email service.

\para{Selecting Target Email Providers.}
This study focuses on popular and public email
services with two considerations. First, popular email services such as Yahoo Mail and Gmail are
used by more than one billion users~\cite{gmail1, yahoo1}. Their
security policies and design choices are likely to impact more people.
Second, to perform end-to-end experiments, we
need to collect data from the receiver end. Public email
services allow us to create an account as the receiver.
Our experiment methodology is applicable to private
email services but requires collaborations from the internal users.

To obtain a list of popular public email services, we refer to Adobe's
leaked user database (152 million email addresses, 9.3 million unique
email domains)~\cite{adobe13}. We ranked the email domains based on popularity, and
manually examined the top 200 domains (counting for 77.7\% of all email
addresses). After merging domains from
the same service ({\em e.g.}, {\tt hotmail.com} and {\tt
  outlook.com}) and excluding services that don't allow us to create an
account, we obtained a short list of 28 email domains. To include
the more recent public email services, we searched in Google and
added 6 more to the list (34 in total). We notice that Google's Gmail
and Inbox have very different email interfaces and we treat them as two different services.

In total, we have 35 popular email services which cover 99.8
million email addresses (65.7\%) in the Adobe database. As an
additional reference, we also analyze the Myspace
database (131.4 million email addresses)~\cite{myspace16}. We find
that 101.8 million email addresses (77.5\%) are from the 35 email services,
confirming their popularity.




\subsection{Controlled Parameters}
To examine how different factors affect the outcome of email spoofing,
we apply different configurations to the experiment. We primarily
focus on parameters of three aspects: the spoofed sender address, email content,
and the receiver's email client (user interface).



\para{Spoofed Sender Address.}
The spoofed sender address is likely to affect the spoofing
result. For example, if the spoofed domain ({\tt A.com}) has a valid
SPF/DKIM record, then the receiver (in theory) is able to detect
spoofing. In addition, {\tt A.com}'s SPF/DMARC policy may also affect
the decision of the receiving email service.
We configure three profiles for the
spoofed sender domain: (1) no SPF/DKIM/DMARC record; (2) SPF/DKIM with a
``none'' policy; and (3) SPF/DKIM with a ``reject'' policy.
 For each profile, we pick a qualified domain for the primary
 experiment ({\tt easychair.org}, {\tt xxx.edu},\footnote{For anonymous submission, we
    use ``{\tt xxx.edu}'' to represent our institution's email
    domain. ``xxx'' does not necessarily mean it contains three letters.}
and {\tt facebook.com}). We use three addresses to intuitively illustrate
the problem. Then to validate the results, we perform a
shadow experiment by randomly selecting 20 domains for each category
(60 in total) from the Alexa top 2000 domains. The shadow domains are
listed in Appendix B.



\para{Email Content.}
Email content can affect how spam filters handle the
email~\cite{Blanzieri2008}. Different keywords may have
different weights for the spam filter, leading to almost an infinite
testing space. Our experiment, however, focuses on spoofing (sender address is
forged) instead of spam, and we want
to minimize the impact of spam
filters. To this end, we configure 3 emails with blank content to measure the
decision made on forgery alone. The email
content is configured as (1) a blank email, (2) a blank email with a benign URL ({\tt
  http://google.com}), and (3) a blank email with a benign attachment
(empty text file). The reason for using ``benign'' content is to test how much the
``spoofing'' factor alone contributes to the email providers' decisions.
In addition, to test whether a phishing email can penetrate the target service, we also include (4) an email with
phishing content. This phishing email is a real-world
sample from a phishing attack targeting our institution in 2017. The
email impersonates the technical support to notify the victim
that her internal account has been suspended and ask her to
re-activate the account using a URL (to an Amazon EC2 server).



\para{Email Client.} We examine how different
email clients warn users of forged emails. We consider 3 common email clients: (1) a web client,
(2) a mobile app, and (3) a third-party email client. All the 35 selected services
have a web interface, and 28 have a dedicated mobile app. Third-party
clients refer to the email applications ({\em e.g.}, Microsoft Outlook
and Apple Mail) that allow users to check
emails from any email providers.

%% file: result.tex
\begin{table*}[t]
\begin{center}
\caption{Results of the end-to-end spoofing experiment (primary). To better illustrate the results, (\inbox) indicates the email is delivered to the inbox; (\spam)
  indicates the email is placed to the spam folder; (\block) indicates the email is
  blocked without delivery. }
\label{tab:bigtable}
\vspace{-0.1in}
\begin{tabu}{r P{0.5cm} P{0.5cm} P{0.795cm} |P{0.56cm} P{0.56cm} P{0.56cm} P{0.56cm} |P{0.56cm} P{0.56cm} P{0.56cm} P{0.56cm}|P{0.56cm} P{0.56cm} P{0.56cm} P{0.56cm}|c}
\tabucline[1.1pt]{-}
Email Provider & \multicolumn{3}{c}{Supported} &
\multicolumn{4}{c}{Spoof {\tt easychair.org}} &
\multicolumn{4}{c}{Spoof {\tt xxx.edu} } &
\multicolumn{4}{c}{Spoof {\tt facebook.com}}
& \# to\\
& \multicolumn{3}{c}{Protocols} &
\multicolumn{4}{c}{(No SPF/DKIM/DMARC)} &
\multicolumn{4}{c}{(SPF/DKIM;DMARC=none)} &
\multicolumn{4}{c}{(SPF/DKIM;DMARC=strict)}
&Inbox
\\
& 	SPF & DKIM & DMARC  & Blank & URL & Attach & Phish & Blank & URL
& Attach & Phish& 	Blank & URL & Attach & Phish
& \\
\taburowcolors{gray!10 .. white}

hotmail.com & \cmark & \cmark & \cmark
& 	\spam & 	\spam & 	\spam & 	\spam
& 	\spam & 	\spam & 	\spam & 	\spam
& 	\block & 	\block & 	\block & 	\block
& 0/12
\\

aol.com & \cmark & \cmark & \cmark
& 	\spam & \spam  & \spam & 	\block
& 	\inbox & 	\spam & 	\spam & 	\block
& \block & \block & \block & \block
& 1/12
\\

seznam.cz & \cmark & \cmark & \cmark
& 	\spam & 	\spam & 	\spam & 	\spam
& 	\spam & 	\spam & 	\spam & 	\inbox
& 	\block & 	\block & 	\block & 	\block
& 1/12\\

163.com & \cmark & \cmark & \cmark
& 	\spam & 	\inbox & 	\spam & 	\spam
& 	\spam & 	\spam & 	\spam & 	\inbox
& 	\spam & 	\spam & 	\spam & 	\block 
& 2/12
\\

126.com & \cmark & \cmark & \cmark
& 	\inbox & 	\inbox & 	\inbox & 	\spam
& 	\spam & 	\spam & 	\spam & 	\spam
& 	\spam & 	\spam & 	\spam & 	\block 
& 3/12
\\

yeah.net & \cmark & \cmark & \cmark
& 	\inbox & 	\inbox & 	\inbox & 	\spam
& 	\spam & 	\spam & 	\spam & 	\spam
& 	\spam & 	\spam & 	\spam & 	\block 
& 3/12
\\

yahoo.com & \cmark & \cmark & \cmark
& 	\spam & 	\spam & 	\spam & 	\spam
& 	\inbox & 	\inbox & 	\inbox & 	\inbox
& 	\block & 	\block & 	\block & 	\block
& 4/12
\\

mail.ru & \cmark & \cmark & \cmark
& 	\spam & 	\spam & 	\inbox & 	\spam
& 	\inbox & 	\inbox & 	\inbox & 	\inbox
& \block  & 	\block & 	\block & 	\block
& 5/12
\\

tutanota.com & \cmark & \cmark & \cmark
& 	\inbox & 	\inbox & 	\spam & 	\inbox
& 	\inbox & 	\inbox & 	\spam & 	\inbox
& 	\block & 	\block & 	\block & 	\block
& 6/12\\

gmail.com & \cmark & \cmark & \cmark
& 	\inbox & 	\inbox & 	\inbox & 	\inbox
& 	\inbox & 	\inbox & 	\inbox & 	\spam
& 	\block & 	\block & 	\block & 	\block
& 7/12\\

gmail inbox & \cmark & \cmark & \cmark
& 	\inbox & 	\inbox & 	\inbox & 	\inbox
& 	\inbox & 	\inbox & 	\inbox & 	\spam
& 	\block & 	\block & 	\block & 	\block
& 7/12\\

icloud.com & \cmark & \cmark & \cmark
& 	\spam & 	\inbox & 	\inbox & 	\spam
& 	\spam & 	\inbox & 	\inbox & 	\inbox
& 	\spam & 	\inbox & 	\inbox & 	\spam
& 7/12\\

naver.com & \cmark & \cmark & \cmark
& 	\inbox & 	\inbox & 	\inbox & 	\inbox
& 	\inbox & 	\inbox & 	\inbox & 	\inbox
& 	\block & 	\block & 	\block & 	\block
& 8/12\\

protonmail.com & \cmark & \cmark & \cmark
& 	\inbox & 	\inbox & 	\inbox & 	\inbox
& 	\inbox & 	\inbox & 	\inbox & 	\inbox
& 	\spam & 	\spam & 	\spam & 	\spam
& 8/12\\

fastmail.com & \cmark & \cmark & \cmark
& 	\inbox & 	\inbox & 	\inbox & 	\inbox
& 	\inbox & 	\inbox & 	\inbox & 	\inbox
& 	\spam & 	\spam & 	\spam & 	\spam
& 8/12\\

inbox.lv &  \cmark & \cmark & \cmark
& \inbox & 	\inbox & 	\inbox & 	\inbox
& \inbox & 	\inbox & 	\inbox & 	\inbox
& \spam & \spam & \spam & \spam
& 8/12\\

\hline

runbox.com & \cmark & \cmark & \xmark
& 	\spam & 	\spam & 	\spam & 	\spam
& 	\spam & 	\spam & 	\block & 	\spam
& 	\spam & 	\spam & 	\block & 	\spam
& 0/12\\

o2.pl & \cmark & \cmark & \xmark
& 	\spam & 	\spam & 	\spam & 	\inbox
& 	\spam & 	\spam & 	\spam & 	\spam
& 	\block & 	\block & 	\block & 	\block
& 1/12\\

wp.pl & \cmark & \cmark & \xmark
& 	\spam & 	\spam & 	\spam & 	\inbox
& 	\spam & 	\spam & 	\spam & 	\spam
& 	\block & 	\block & 	\block & 	\block
& 1/12\\

interia.pl & \cmark & \xmark & \xmark
& 	\inbox & 	\spam & 	\inbox & 	\inbox
& 	\inbox & 	\spam & 	\inbox & 	\spam
& 	\block & 	\block & 	\block & 	\block
& 5/12\\

sapo.pt & \cmark & \xmark & \xmark
& 	\inbox & 	\inbox & 	\spam & 	\inbox
& 	\spam & 	\inbox & 	\spam & 	\inbox
& 	\block & 	\block & 	\block & 	\block
& 5/12\\

mynet.com & \cmark & \cmark & \xmark
& 	\inbox & 	\inbox & 
\block  & 	\spam
& 	\spam & 	\spam & 
\block  & 	\spam
& 	\inbox & 	\inbox & 	\inbox & 	\inbox
& 6/12\\

op.pl & \cmark & \cmark & \xmark
& 	\inbox & 	\inbox & 	\inbox & 	\inbox
& 	\inbox & 	\inbox & 	\inbox & 	\inbox
& 	\block & 	\block & 	\block & 	\block
& 8/12\\

gmx.com  & \cmark & \cmark & \xmark
& 	\inbox & 	\inbox & 	\inbox & 	\inbox
& 	\inbox & 	\inbox & 	\inbox & 	\inbox
& 	\spam & \spam & \spam &	\spam
& 8/12\\

mail.com & \cmark & \cmark & \xmark
& 	\inbox & 	\inbox & 	\inbox & 	\inbox
& 	\inbox & 	\inbox & 	\inbox & 	\inbox
& 	\spam & \spam & \spam & \spam
& 8/12\\

qq.com & \cmark & \cmark & \xmark
& 	\inbox & 	\inbox & 	\inbox & 	\inbox
& 	\spam & 	\inbox & 	\spam & 	\inbox
& 	\inbox & 	\inbox & 	\inbox & 	\inbox
& 10/12\\

daum.net & \cmark & \xmark & \xmark
& 	\inbox & 	\inbox & 	\inbox & 	\inbox
& 	\inbox & 	\inbox & 	\inbox & 	\spam
& 	\inbox & 	\inbox & 	\inbox & 	\block
& 10/12\\

zoho.com & \cmark & \cmark & \xmark
& 	\inbox & 	\inbox & 	\inbox & 	\inbox
& 	\inbox & 	\inbox & 	\inbox & 	\inbox
& 	\inbox & 	\spam & 	\inbox & 	\inbox
&11/12\\

sina.com & \cmark & \cmark & \xmark
& 	\inbox & 	\inbox & 	\inbox & 	\inbox
& 	\inbox & 	\inbox & 	\inbox & 	\inbox
& 	\inbox & 	\inbox & 	\inbox & 	\inbox
&12/12\\

juno.com & \cmark & \cmark & \xmark
& 	\inbox & 	\inbox & 	\inbox & 	\inbox
& 	\inbox & 	\inbox & 	\inbox & 	\inbox
& 	\inbox & 	\inbox & 	\inbox & 	\inbox
&12/12\\

sohu.com & \cmark & \xmark & \xmark
& 	\inbox & 	\inbox & 	\inbox & \inbox
& 	\inbox & 	\inbox & 	\inbox & \inbox
& 	\inbox & 	\inbox & 	\inbox & \inbox
& 12/12\\

\hline

rediffmail.com & \xmark & \xmark & \xmark
& 	\inbox & 	\inbox & 	\inbox & 	\inbox
& 	\inbox & 	\inbox & 	\spam & 	\spam
& 	\inbox & 	\spam & 	\spam & 	\spam
& 7/12\\

freemail.hu & \xmark & \xmark & \xmark
& 	\inbox & 	\inbox & 	\inbox & \inbox
& 	\inbox & 	\inbox & 	\block & \inbox
& 	\inbox & 	\block & 	\block & \spam
& 8/12\\

t-online.de & \xmark & \xmark & \xmark
& 	\inbox & 	\inbox & 	\inbox & 	\inbox
& 	\inbox & 	\inbox & 	\inbox & 	\inbox
& 	\inbox & 	\inbox & 	\block & 	\inbox
& 11/12\\

excite.com & \xmark & \xmark & \xmark
& 	\inbox & 	\inbox & 	\inbox & 	\inbox
& 	\inbox & 	\inbox & 	\inbox & 	\inbox
& 	\inbox & 	\inbox & 	\inbox & 	\inbox
& 12/12\\

\tabucline[1.1pt]{-}
\end{tabu}

\end{center}
\vspace{-0.2in}
\end{table*}

\section{Spoofing Experiment Results}
\label{sec:result}

In this section, we describe our findings from the end-to-end spoofing
experiments. First, to provide the context, we measure the authentication
mechanisms (SPF/DKIM/DMARC) that the email providers use to detect
forged emails. Then, we examine how email providers handle forged
emails and identify key factors that help forged emails to reach user
inbox. For emails that successfully get in, we
examine whether and how email providers warn users about their
potential risks. Finally, we present case studies on email
providers that have misleading UIs which help the
attacker to make the forged email look more authentic.




\subsection{Authentication Mechanisms}
To better interpret the results, we first examine how the
35 email providers authenticate incoming emails. One
way of knowing their authentication protocols is to analyze the email
headers and look for authentication results of SPF/DKIM/DMARC. While
this method works for some of the email providers, it will miss those ({\em
  e.g.}, {\tt qq.com}) that do not add the authentication results to
the header. Instead, we follow a more reliable
measurement method~\cite{Foster:2015}. The high-level idea is to setup
an {\em authoritative} DNS server for
our own domain. Then we send a forged
email to the target email service by spoofing our own domain. In the
meantime, the authoritative DNS server will wait and see whether the target
email service will query our SPF/DKIM/DMARC record. We set the TTL of
the SPF, DKIM and DMARC records as 1 (second) to force the target
email service always querying our {\em authoritative} DNS server for
these records. The results (collected in August 2017) are shown in
Table~\ref{tab:bigtable} (left 4 columns). 35 email providers can be
grouped into 3 categories based on their protocols:

\begin{packed_itemize}
\item {\bf Full Authentication (16):} Email services that perform all three authentication checks (SPF, DKIM and
DMARC). This category includes the most popular email services such as
Gmail, Yahoo Mail, Hotmail and Apple's iCloud.

\item {\bf SPF/DKIM but no DMARC (15):} Email services that check either SPF/DKIM to
authenticate incoming emails, but do not check the sender's DMARC
policy. These email services are likely to make decisions on their own.

\item {\bf No Authentication (4):} Email services that do not perform any of
the three authentication protocols.
\end{packed_itemize}
During this experiment, we observe that {\tt gmx.com}, {\tt mail.com}
and {\tt inbox.lv} block all the emails from dynamic IPs. In our
spoofing experiment, we use dynamic an IP as default to simulate a
worse-case scenario for attackers, but we use a static IP for these three
services.






\subsection{Decisions on Forged Emails}
Next, we examine the decision-making process on forged emails. As a
primary experiment, we send 12 forged emails to each of the 35 email
services (3 spoofed addresses $\times$ 4 types of email content). In
August 2017, we shuffled all
the emails and set a sending time interval of 10 minutes to minimize
the impact to the target services. Note that we do not send a large
volume of emails (only 12 emails per service). This traffic volume is
extremely low compared to the {\em hundreds of billions} of emails sent over
the Internet every day~\cite{email-s17}. We intentionally limit our
experiment scale so that the experiment emails would not impact the
target services (and their email filters) in any significant
ways. This is also to avoid emails sent earlier affecting the ones
sent later in the experiments. To validate
the results of the primary experiment, we will also conduct a shadow
experiment by testing more spoofed domains.

The experiment results are shown in Table~\ref{tab:bigtable}. We group
email providers based on the supported authentication protocols. Then
within each category, we rank email providers based on the number of
forged emails that arrived the inbox.


\para{Observations.} 
First, we find that forged emails are able to penetrate
most of the email services. 33 out of the 35 services allowed at least one
forged email to arrive the inbox. 30 services allowed at least
one {\em phishing} email to get into the inbox. Particularly, this
phishing email has penetrated email providers that perform full
authentications ({\em e.g.}, Gmail, iCloud, Yahoo Mail) when spoofing
{\tt easychair.org} and {\tt xxx.edu}. This suggests that even though
the email providers detected the email forgery (based on SPF/DKIM),
they still deliver the phishing message to the user inbox.

\begin{figure*}
\centering
    \subfigure[Primary Experiment (All)]{
      \includegraphics[width=0.3\textwidth]{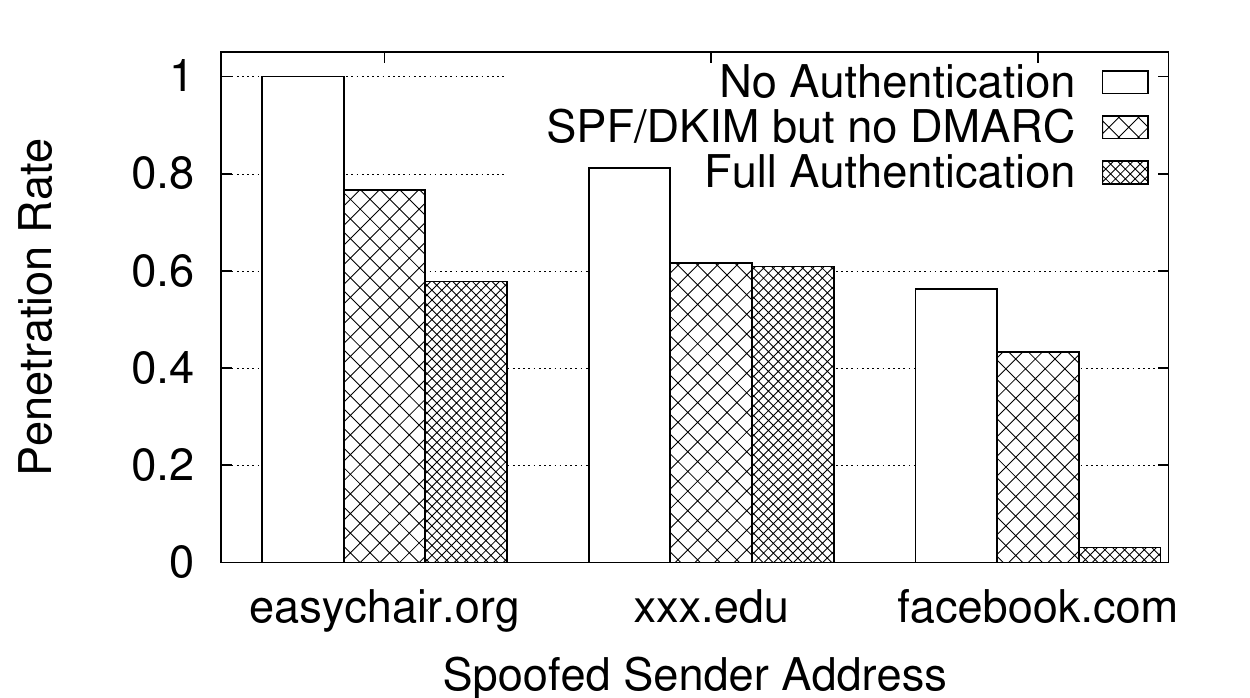}
       \vspace{-0.05in}
      \label{fig:blocking}
      \vspace{-0.05in}
    }
    \hfill
    \subfigure[Primary Experiment (Phishing Content) ]{
      \includegraphics[width=0.3\textwidth]{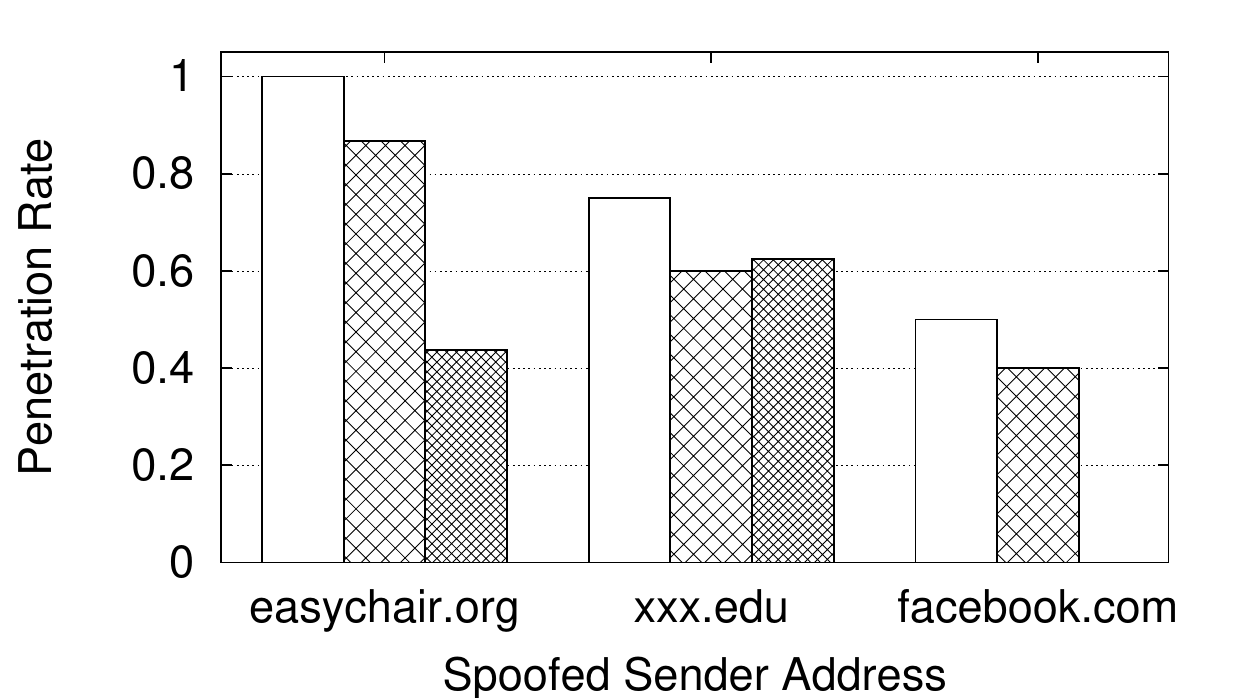}
     \vspace{-0.05in}
      \label{fig:blocking_p}
      \vspace{-0.05in}
    }
    \hfill
    \subfigure[Shadow Experiment (Phishing Content)]{
      \includegraphics[width=0.3\textwidth]{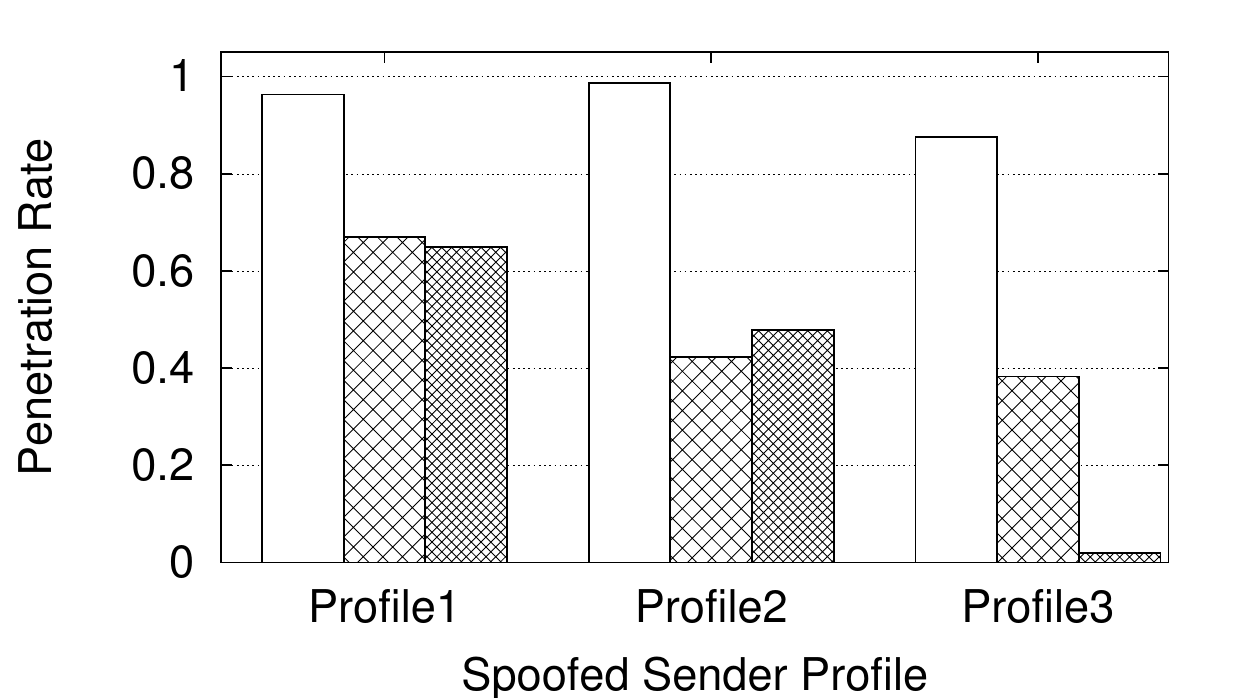}
     \vspace{-0.05in}
      \label{fig:shadow}
      \vspace{-0.05in}
    }
\vspace{-0.08in}
  \caption{Penetration rate for different combinations of the sender
    and the receiver. The x-axis shows 3 spoofed sender profiles:
Profile1: No SPF/DKIM/DMARC (easychair.org); Profile2: SPF/DKIM and
none policy (xxx.edu); Profile 3: SPF/DKIM and reject policy
(facebook.com). The legend displays the 3 authentication
groups of the receivers. The primary experiment covers 3 sender
addresses. The shadow experiment covers 60 sender addresses, 20
for each sender profile. }
\vspace{-0.1in}
  \label{fig:allexp}
\end{figure*}

Second, comparing different email providers, we observe that those
without authentication schemes are indeed more vulnerable. To better
illustrate the trend, we plot Figure~\ref{fig:blocking} and
Figure~\ref{fig:blocking_p} to show the penetration rate (the
ratio of emails in the inbox over all the emails). 
Clearly, email providers with no authentications have the highest penetration rate ranging from
0.56--1.0 depending on the spoofed address.
Since these services did not check the SPF/DKIM
record, it is possible they rely on the domain reputation or
sensitivity to take different actions ({\em
  e.g.}, Facebook is more sensitive than Easychair).
As a comparison, email providers that perform authentications have a lower penetration
rate: 0.43--0.77 for those that only perform SPF/DKIM, and 0.03--0.61
for those that perform full authentications.

Third, from the sender domain's perspective, setting a
``reject'' policy makes a big difference. For example, {\tt facebook.com}
has a ``reject'' DMARC policy. As long as the email providers check
the DMARC record, almost 100\% of them rejected the emails. The only exception is iCloud which allows two forged
Facebook emails to pass. We suspect that iCloud treated the sender policy as a
``feature'' in their decision function instead of enforcing the
policy.



\para{Shadow Experiment.}
While the primary experiments help to illustrate the problem in an
intuitive manner, the choice of the 3 spoofed domains may
introduce biases. Here we run a shadow experiment to make sure our
observations are not entirely dependent on the
3 selected sender addresses. As previously described, the shadow
experiment spoofs 60 domains, 20 for each of the 3 sender profiles: (1) No
SPF/DKIM/DMARC; (2) SPF/DKIM and ``none'' policy; (3) SPF/DKIM and ``reject''
policy. The complete list of sender domains is shown in Appendix B. 
The emails are also shuffled and sent with a 10-minute interval. 
The result is shown in Figure~\ref{fig:shadow}. The penetration rates are not
exactly the same with the primary experiment but the overall trends and
conclusions remain consistent. For example, {\tt
  facebook.com} has a lower penetration rate than the corresponding
group in the shadow experiment. This is likely because {\tt
  facebook.com} is a more ``sensitive'' domain. The consistent trends
are: (1) domains that published SPF/DKIM/DMARC
and/or strict policies are harder to spoof; (2) email providers that
perform authentication are harder to penetrate.



\para{Impacting Factors in the Experiment.}
To determine which factors contribute more to a successful
penetration, we perform a ``feature ranking'' analysis. We divide all the emails in the primary and shadow
experiments into two classes: {\em
  positive} (inbox) and {\em negative} (spam folder or blocked). 
For each email, we calculate three features: email
content ($F_1$), sender profile ($F_2$) and receiver authentication
group ($F_3$), all of which are categorical variables. Then we rank
features based on their distinguishing power to classify emails into the
two classes. We use two widely used feature ranking metrics: Chi-Square
Statistics~\cite{lancaster1969chi} and Mutual
Information~\cite{cover2012elements}. As shown in
Table~\ref{tab:rank}, consistently, ``spoofed sender profile'' is the most important factor, followed by
the ``receiver authentication method''.  
Note that this analysis only compares the relative importance of factors in our
experiment. We are not trying to reverse-engineer the whole spoofing
detection system, which will require analyzing more features. 

\begin{table}[t]
\centering
\caption{Feature Ranking.}
\vspace{-0.05in}
\label{tab:rank}
\begin{tabu}{l|l|l}
\tabucline[1.1pt]{-}
Feature & Chi-square & Mutual Information \\\hline
Spoofed sender profile & 202.30 & 0.063 \\
Receiver authentication method & 152.59 &  0.058 \\
Email content & 0.62 & 0.00069 \\
\tabucline[1.1pt]{-}
\end{tabu}
 \vspace{-0.1in}
\end{table}



\para{Discussion.} 
It takes both the sender and the receiver email services to make a
reliable email authentication. When one of them fails to do their job,
there is a higher chance for the forged email to get into the user
inbox. In addition, we find that email providers have the tendency to prioritize
email delivery over security. When an email fails the SPF/DKIM/DMARC check,
most of the providers (including Gmail and iCloud) would still deliver the
email as long as the policy of the spoofed domain is not
``reject''. Based on the earlier measurement result (\S\ref{sec:1m}),
only 12.3\% of the top 1 million domains have set a ``reject'' or
``hard fail'' policy, which leaves plenty of room for attackers to perform
spoofing attacks.

\begin{figure*}
         \begin{center}
      	\begin{minipage}[t]{0.99\textwidth}
   	\includegraphics[width=0.99\textwidth]{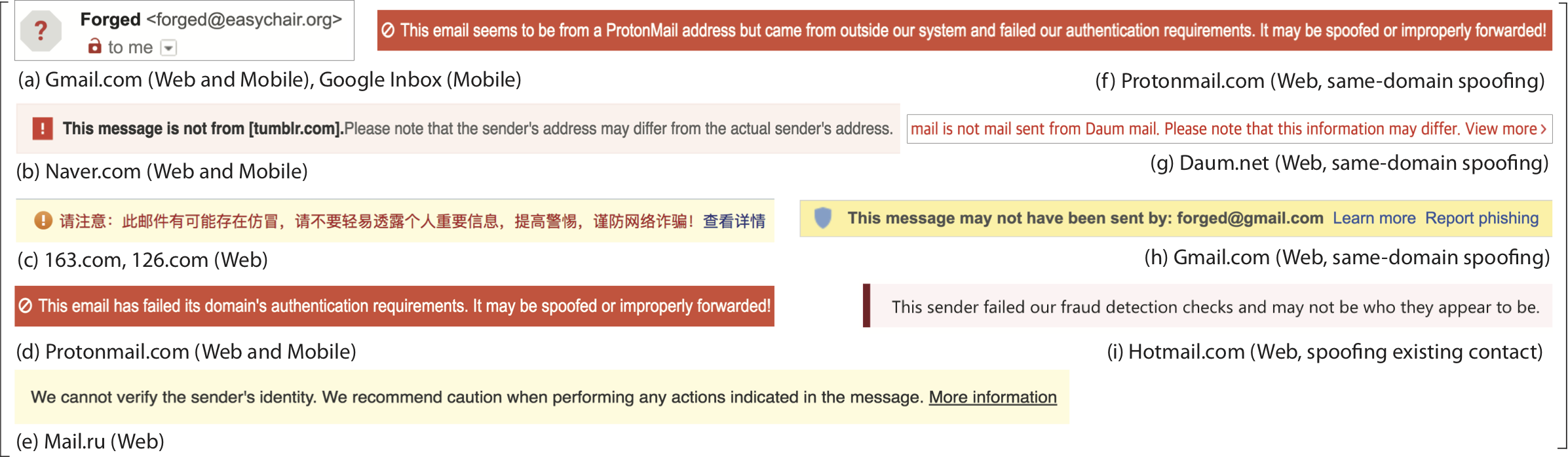}
	\end{minipage}
\caption{Security cues on forged
  emails from 9 email providers. (a)--(e) are for regular forged emails. (f)--(h) only show up when the spoofed sender
  and the receiver belong to the same provider. (i) only shows up when spoofing
  an existing contact.}
	\label{fig:screen}
  	\vspace{-0.12in}
\end{center}
\end{figure*}



\subsection{Email Clients and Security Cues}
Our results show that most email services allow forged emails to
reach the inbox. Next, we examine email interfaces to see what type
of security cues are displayed to warn users.

\para{Web Client.} During our experiment, only 6 email services have
displayed security cues on forged emails to alert users, including
{\tt gmail}, and {\tt protonmail} (US),  {\tt naver}
(South Korea), {\tt mail.ru} (Russia), and 2 services from the same company in China: {\tt
  163.com} and {\tt 126.com}. 
All the other email services display forged emails without any visual alert
(\eg, Yahoo Mail, iCloud). Note that Gmail and Google Inbox are 
developed by the same company, but the web version of Google Inbox has no security cue.

Figure~\ref{fig:screen} (a)--(e) show the screenshots of the security
cues. Gmail's cue is a ``question mark'' on the sender's icon. Only
when users move the mouse over the image, it will show the following
message: {\em ``Gmail could not verify that $<$sender$>$ actually sent this
message (and not a spammer)''}. The red lock icon is not related to
spoofing, but to indicate the communication between MX
servers is unencrypted. On the other hand, services like {\tt naver}, {\tt 163.com}
and {\tt protonmail} use explicit text messages to warn users.


\para{Mobile Client.} Compared to the web interface, mobile apps have even
fewer security cues. Out of the 28 email services with
a dedicated mobile app, only 4 services have mobile security cues
including {\tt naver}, {\tt protonmail}, {\tt gmail}, and {\tt google inbox}. The other services removed the security
cues for mobile users. Compared to the web interface, mobile apps
have more limited screen size. Developers often remove or hide less important
information and UI elements to keep a ``clean''
interface. Unfortunately, the security cues are among the removed
elements.

\para{Third-party Client.} Finally, we check emails using third-party
clients including Microsoft Outlook, Apple Mail, and Yahoo
Web Mail. We test both desktop and mobile
versions, and find that {\em
  none} of them provide security cues for forged emails. This is not too surprising
considering that the email client and the mail server are from different
companies. 




\subsection{Misleading UI Elements}
Under special conditions, attackers may trigger
misleading elements on the user interface (UI) to make the forged
email more realistic. In the following, we describe three special
conditions that can trigger misleading UIs.

\para{Spoofing an Existing Contact. }
We find that when an attacker spoofs an existing contact of the receiver, the
forged email can trigger misleading UI elements. Certain email
services will automatically load the contact's photo, name card or previous email conversations
alongside the forged email, which helps the attacker to make
the forged email look authentic. To demonstrate this phenomenon, we perform a quick
experiment as follows: First, we create an ``existing contact'' ({\tt contact@xxx.edu}) for each
receiver account in the 35 email services, and add a name, a profile
photo and a phone number (if allowed). Then we spoof
this contact's address ({\tt contact@xxx.edu}) to send 4 forged
emails to each target service (4 types of content).

Table~\ref{tab:contact} shows the 25 email providers
that have misleading UIs: 6 services automatically load the contact's
profile photo, 17 services have a clickable name card, and 17 services
display a link (or a widget) for historical emails with this contact (screenshots in
Appendix C). We believe that these designs aim to improve the usability of the
email service by providing the context information for the sender.
However, when the sender address is actually
spoofed, these UI elements would help attackers to make the
forged email look more authentic.

In addition, spoofing an existing contact allows forged emails to
penetrate new email providers. For example, Hotmail blocked {\em all} the
forged emails in Table~\ref{tab:bigtable}. However, when we spoof an
existing contact, Hotmail delivers the forged email to the inbox and
adds a special warning sign as shown in Figure~\ref{fig:screen}(i).

\begin{table}[t]
\centering
\caption{Misleading UI elements when the
  attacker spoofs an existing contact.  ($*$) indicates web interface only. ($\dag$)
  indicates mobile only. Otherwise, it applies to both interfaces. }
\label{tab:contact}
\vspace{-0.05in}
\begin{tabu}{l|l}
\tabucline[1.1pt]{-}
Misleading UI & Email Providers (25 out of 35)\\     \hline
Sender Photo (6)  &
G-Inbox,
gmail,
zoho,
icloud$^*$, 
gmx$^\dag$,
mail.com$^\dag$
\\
\hline
Name Card (17)   &
yahoo,
hotmail,
tutanota,
seznam.cz,
fastmail, 
gmx,
\\ &
mail.com, 
gmail$^*$,
sina$^*$,
juno$^*$,
aol$^*$,
163.com$^\dag$,
\\
&
126.com$^\dag$,
yeah.net$^\dag$,
sohu$^\dag$,
naver$^\dag$,
zoho$^\dag$
\\
\hline
Email History (17)  &
hotmail,
163.com,
126.com,
yeah.net,
qq, 
zoho,
\\
&
mail.ru,
yahoo$^*$,
gmail$^*$,
sina$^*$,
naver$^*$, 
op.pl$^*$,
\\
&
interia.pl$^*$,
daum.net$^*$
gmx.com$^*$, 
mail$^*$,
inbox.lv$^*$
\\
\tabucline[1.1pt]{-}
\end{tabu}
\end{table}

\begin{table}[t]
\centering
\caption{Misleading UI elements during
  the same-domain spoofing. ($\dag$)
  indicates mobile interface only.}
\label{tab:same}
\vspace{-0.05in}
\begin{tabu} {p{2.1cm}|p{5.55cm}}
\tabucline[1.1pt]{-}
Misleading UI & Email Providers (3 out of 35)\\     \hline
Sender Photo (3) &
seznam.cz,
google Inbox,
gmail$^\dag$
\\
\tabucline[1.1pt]{-}
\end{tabu}
\vspace{-0.1in}
\end{table}

\begin{figure}[t]
\centering
\begin{minipage}[t]{0.48\textwidth}
\includegraphics[width=0.8\textwidth]{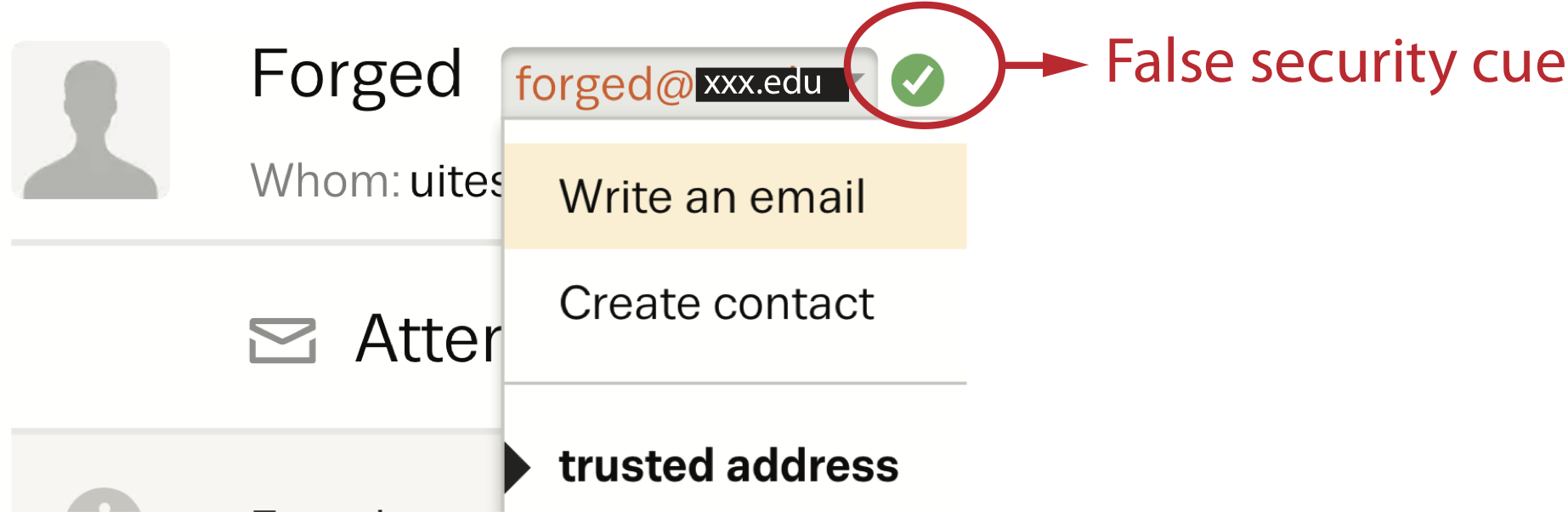}
 \vspace{-0.07in}
\caption{Seznam.cz displays a ``trusted address'' sign on a forged
  address.}
\label{ui:seznamczbug}
  	\vspace{-0.18in}
\end{minipage}
\end{figure}



\para{Same-domain Spoofing.}
Another way to trigger the misleading UI element is to spoof an email
address that belongs to the same email provider as the receiver (Table~\ref{tab:same}). For
example, when spoofing {\tt <forged@seznam.cz>} to send an email to
{\tt <test@eznam.cz>}, the profile photo of the spoofed sender will be
automatically loaded onto the interface. Because the spoofed sender is
also from the same provider, the email provider can easily load the sender's
photo from its own database. This phenomenon applies to Google Inbox
and Gmail (mobile) too.


Same-domain spoofing, however, is more difficult to deliver the forged
email to the inbox. As a simple experiment, we send 4 emails each
target email provider by spoofing their own domains. 
  Only 14 email providers have allowed a forged email to reach
the inbox (the corresponding number is 33 for Table~\ref{tab:bigtable}). 
Providers also alert users more with
special security cues. As shown in Figure~\ref{fig:screen}(f)-(h),
related email providers include {\tt protonmail}, {\tt gmail} and {\tt
  daum.net}. This suggests that email providers are doing a better job in
detecting forged emails that claim to be from their own services.
Together with previously observed security cues, there are in total 9 email services
that provide at least one type of security cues.

\para{False Security Cues.}
One email provider {\tt seznam.cz} displays a false
security cue to users. {\tt seznam.cz} performs full authentications
but still delivers a phishing email to the inbox. Figure~\ref{ui:seznamczbug} shows that
{\tt seznam.cz} displays a green checkmark on the sender address even though the
address is forged. When users click on the icon, it displays
``trusted address'', which is likely to give users a false sense of security.


\subsection{Other Vulnerabilities}
We find that 2 email services ``{\tt sapo.pt}''
and ``{\tt runbox.com}'' are not carefully configured, allowing an
attacker to piggyback on their mail servers to send forge emails. This threat model is very different from our experiments
above, and we briefly describe it using Figure~\ref{fig:sys}. Here, the attacker is the sender MUA, and the vulnerable server
({\em e.g.}, {\tt runbox.com}) is the sender service. Typically,
Runbox should only allow its users to send an email with
the sender address as ``{\tt \{someone\}@runbox.com}''. However, the
Runbox's server allows a user (the attacker) to set the ``{\tt
  MAIL FROM}'' freely (without requiring a verification) in step \ding{182} to send forged
emails. This attack does not help the forged email to bypass the
SPF/DKIM check. However, it gives the attacker a {\em static and reputable}
IP address. If the attacker aggressively sends malicious emails
through the vulnerable mail server, it can damage the reputation of
the IP. We have reported the vulnerability to the service admins.

%% file: user.tex
\section{Effectiveness of Security Cues}
\label{sec:design}
Thus far, our results show that email providers tend to prioritize
email delivery over security when the sender address cannot be
verified. As a result, forged emails have the chance to reach the user
inbox. While most email services fail to warn users, a few did have
implemented visual security cues on the user interface. In the
following, we seek to understand how effective these security cues
are to improve user efficacy in detecting phishing emails. 


\subsection{Challenges in Security Usability Study}
To evaluate the effectiveness of security cues, we seek to design
a user study where participants examine phishing
emails (with forged sender addresses). By controlling the security cues on
the interface, we assess how well security cues help users to handle
phishing emails more securely.

Implementing this idea faces a key challenge in security
usability study, which is to capture realistic user reactions to
security attacks. Ideally, participants should
examine phishing emails {\em without knowing that they are in
  an experiment}. This leads to key challenges to: (1) to
set up the experiment and obtain user consent
upfront; (2) to test a variety
of experiment conditions since each participant can only be
``deceived'' once.

As a result, most security usability studies follow a ``role-playing'' approach,
where participants are instructed to play a pre-defined role in a
hypothetical scenario~\cite{jJPMB12, Sheng:2010:, Vishwanath:2011, Wang:2016, Wu:2006:}. 
Researchers often hide the true purpose of the study and observe the user behavior.
The ``role-playing'' method allows researchers to test a variety of
controlled settings. The drawback, however, is
that participants may behave differently from the real-world.

To this end, we explore a hybrid approach by combining role-playing
experiments with real-world deceptive studies
(Table~\ref{tab:exp1}). We leverage the controlled {\em role-playing experiments} to test a
variety of conditions and then use a {\em deceptive phishing
  experiment} to perform a more focused study. 


\begin{table}[t]
\centering
\caption{Experiment settings.}
\label{tab:exp1}
\vspace{-0.1in}
\begin{tabu}{lll}
\tabucline[1.1pt]{-}
{\bf Experiment ($N=913$)} & {\bf Conditions} & {\bf Variables} \\
\hline
Study 1 ($N_1=425$) & role-playing & content (4) $\times$ security cue (2) \\
\hline
Study 2 ($N_2=488$) & real-world & content (1) $\times$ security cue (2)\\
\tabucline[1.1pt]{-}
\end{tabu}
\vspace{-0.23in}
\end{table}


\subsection{Study 1: Role-Playing Experiment}
Fist, we conduct a ``role-playing'' experiment where participants play a pre-defined role to
read a set of phishing and non-phishing emails.


\para{Study Procedure.}
We conduct a user survey where the participant plays the role of an
employee (named Pat Jones) in a company. The participant
is expected to see emails from his colleagues, the company's IT department, and
various online services. Each email is presented as a {\em screenshot} of the
full email interface. After the participant reads each email, we ask
``how would you respond to this email'' with 8 options including
``Reply by email'', ``Contact the sender by phone or in person'', 
``Delete the email'', ``Do nothing'', ``Click on the URL'',  ``Copy
and paste the URL to the browser'', ``Visit the website
directly'', and  ``Others (please specify)''. The last option is to
collect open answers. The participant can check multiple options that apply. 

Like most role-playing studies, we avoid mentioning the keywords
``security'' or ``phishing'' in the instruction. Instead, we distract the
participants using a non-security task, stating that the
survey is to study email usage habit, email miscommunications
and how people handle different types of emails. 
We also ask a few distraction questions regarding
how long they have been using email
services, and how often they check their personal emails.

\begin{table}[t]
\caption{Email samples used in our user study, including 4 phishing
  emails (P) and 4 legitimate emails (Legit). }
\label{tab:sample}
\vspace{-0.15in}
\begin{center}
\begin{tabu}{l|l}
\tabucline[1.1pt]{-}
{\bf Email ID} & {\bf A Brief Description of the Email} \\
\hline
P-Greed & Upgrading your Office 360 for free.  \\
P-Curiosity & Your message delivery is delayed, check the reason here. \\
P-Urgency & You got 24 hours to verify your account.\\
P-Fear & Our evidence shows that you sent a terroristic threat. \\
\hline
Legit-1 & An updated meeting schedule sent from a colleague. \\
Legit-2 & A welcome message from Twitter.  \\
Legit-3 & Try-one-month-for-free from Netflix. \\
Legit-4 & Email from Facebook to reset your password.\\
\tabucline[1.1pt]{-}
\end{tabu}
\end{center}
\vspace{-0.2in}
\end{table}





\para{Email Samples.} We select 4 phishing emails representing 4
types of human emotions that are commonly exploited in social
engineering including greed, curiosity, urgency,
and fear~\cite{macfee}. Table~\ref{tab:sample} shows a brief description
of the email content. All 4 emails are real-world phishing
emails that once targeted our institution. For comparison,
we select another 4 real-world legitimate emails. All
emails contain a URL link in the email body.
To avoid overwhelming the participant, each
participant will only read 5 emails including the 4 legitimate
emails and 1 phishing email.



\para{Security Cues.} Based on our previous measurement, most
email services adopted text-based cues
(Figure~\ref{fig:screen}(b)-(i)). Even Gmail's special cue
(Figure~\ref{fig:screen}(a)) will display a text message when
users move the mouse over. To this end, we use text-based cue and make
two settings, namely {\em with security cue} and {\em without security cue}. We
added the security cue to all the phishing emails' interfaces to
measure its impact. We present the screenshots of the full email
interfaces as shown in Appendix D. We use Yahoo Mail interface to be
consistent with the later {\em study 2}. We also add
the cue to 1 of the 4 legitimate emails to measure the impact of the
false alert.





\para{Recruiting Participants.} 
We recruited $N=425$
participants in total from Amazon Mechanical Turk (MTurk). 
MTurk users usually come from diverse backgrounds, more diverse
than typical college student samples~\cite{Antin:2012:}.
The participants are evenly divided into 8 groups (46-65 users
per group). Each group works under one experiment
setting (8 settings in total: 4 phishing emails $\times$ 2 security
cues). To avoid
non-serious users, we apply the screening criteria that are commonly used in
MTurk~\cite{ndss15:,Gadiraju:2015:}. We recruit users from the
U.S. who have a minimum Human Intelligence Task (HIT) approval rate of
90\%, and more than 50 approved HITs. Each participant can only take
the survey once to earn \$0.5. To make each experiment independent, we
strictly limit a user to only participating in one experiment setting
(the uniqueness is guaranteed both within HITs and across HITs).


Demographic analysis shows that our
participants come from relatively diverse backgrounds: 
44.2\% are male and 55.5\% are female (0.2\% chose not to
disclose). Most participants are 30--39 years old (34.4\%), followed
by users within 19--29 (32.2\%), above 50 (17.2\%) and 40--49
(16.2\%). Most of the participants have a bachelor degree (39.1\%) or a college
degree (29.2\%). About 20.8\% have a graduate degree and 11.1\% are
high-school graduates.

\subsection{Study 2: Real-World Phishing Tests}
Even though the role-playing experiment distracts the participants
with non-security tasks, the experiment design may still affect user behavior. For
example, participants in {\em study 1} receive instructions
to examine emails one by one. The setup is likely to help
participants to be mentally concentrated when reading emails ({\em
  e.g.}, leading to a better performance of detecting spoofing). In
practice, users are usually caught off guard by phishing emails. To these
ends, we design a deceptive experiment to examine a user's
{\em natural reaction} to phishing emails. We follow the common practices of
deceptive studies and obtains informed user consent {\em after} the
experiment.



\para{Study Procedure.} This study contains two phases. Phase1 is to
set up the deception and phase 2 carries out the phishing experiment.
Like {\em study 1}, we also frame the study with a non-security purpose.

{\em Phase1}: The participant starts by entering her {\em own
email address}. Then we immediately send the participant an email and
instruct the participant to check this email from her email
account. The email contains a tracking pixel (a
1$\times$1 transparent image) to measure
if the email has been opened. After that, we ask a few questions about
the email (to make sure they actually opened the email), and other
distractive questions such as their email usage habits. {\em Phase1}
has three purposes: (1) to make sure the participants actually
own the email address they entered; (2) to test if the tracking pixel works, considering that some users
may configure their email service to block images or HTML; (3) to set
up the deception: after phase1, we give the participants the impression
that the survey is completed (participants get paid after
{\em phase1}). In this way, participants would not
expect a second phishing email.



{\em Phase2:} We wait for 10 days and send the second phishing
email. The second email contains a benign URL pointing to our own
server to measure whether the URL is clicked. In addition, the email body contains
a tracking pixel to measure if the email has been opened. The second
email is only sent to users whose email service is not configured to
block HTML or tracking pixels (based
on {\em phase1}). We wait for another 20 days to
monitor user clicks. After the study, we send a debriefing email which
explains the true purpose of the experiment and obtain the
informed consent. Participants can
withdraw their data anytime. By the time of our submission, none of
the users have requested to withdraw their data. 





\para{Email Samples.}
For phase 2, we use {\tt P-Curiosity}, since it
is the most effective content in {\em study 1}. We
impersonate Amazon Mechanical Turk ({\tt support@mturk.com}) to send this email.
For phase 1, we use a different content ({\tt P-Greedy}) to reduce the probability that the
participant realizes the second email is part of the study.

\para{Security Visual Cues.}
To be consistent with  {\em study 1}, we also have 2 settings for
security cues. The key difference is that we no longer use screenshots, but
display the security cues on the actual email 
interface. For the group {\em without security cue}, we recruit users
from Yahoo Mail. We choose Yahoo Mail users because Yahoo Mail is the largest email service
that has not implemented any security cues. For the
comparison group {\em with security
  cue}, we still recruit Yahoo Mail users and add our own security
cues to their interfaces. More specifically, when sending emails, we
can embed a piece of HTML code in the email body mimicking a text-based
cue (see Appendix D). Note that this is how most email providers insert their
warning cues in the email body (except Gmail). To be consistent, {\em
  study 1} was also using Yahoo
Mail's interface to take the screenshots. 




\para{Uncontrolled Parameters.}
Since we cannot give any instructions in {\em phase2}, we cannot control if
a user uses the mobile app or the website to read the email. This is
not a big issue for Yahoo Mail users. Yahoo's web and mobile clients
both render HTML by default. The text-based cue is embedded in the email body by
us, which will be displayed consistently for both web and mobile users. 

\para{Recruiting Participants.} To collect enough data points from {\em
  phase 2}, we need to recruit a large number of users given that many
users may not open our email. In total, we recruited $N=488$ users
from MTurk: 243 users for the ``without security cue'' setting
 and 245 users for the ``with security cue'' setting. Each user can only
participate one setting for only once to receive \$0.5. If a user
already participated in {\em study 1}, this user will
be no longer qualified for {\em study 2}. 
In the recruiting letter, we explicitly informed the users that we
need to collect their email address. This may introduce self-selection
bias: we are likely to recruit people who are willing to
share their email address with our research team. Our result shows
that the resulting user demographics are still very similar to those in {\em
  study 1}: 49\% are male and 51\% are female. Most participants
are 30--39 years old (39.1\%), followed by users under 29
(31.8\%), above 50 (14.5\%), and 40--49 (14.5\%). Most of the
participants have a bachelor degree (35.0\%) or a college degree
(33.8\%), followed by those with a graduate degree (20.7\%) and
high-school graduates (10.5\%).

\subsection{Ethic Guidelines}
\label{sec:ethic2}
Our study received IRB approval, and we have taken active
steps to protect the participants. In {\em study 1}, phishing
emails are presented as screenshots (not clickable) which incur no
risks. For {\em study 2}, only
benign URLs are placed in the emails which point to our own
server. Clicking on the URL does not introduce practical risks to the
participants or their computers. Although we can see the participant's
IP, we choose not to store the IP information in our dataset. In
addition, we followed the recommended practice from IRB to conduct the
deceptive experiment. In the experiment instruction, we omit
information only if it is absolutely necessary ({\em e.g.}, the
purpose of the study and details about the second email). Revealing such
information upfront will invalidate our results.
After the experiment, we immediately contact the participants to
explain our real purpose and the detailed procedure.
We offer the opportunity for the participants to opt out. Users who
opt-out still get the full payment.

\section{User Study Results}
\label{sec:user}
We now analyze the user study results to
answer the following questions. First, how effective are security cues
in protecting users? Second, how does the impact of security
cues vary across different user demographics?



\begin{table}[t]
\caption{Top 3 user reactions without security cues (S1). }
\label{tab:react}
\vspace{-0.15in}
\begin{center}
\begin{tabu}{l|lll}
\tabucline[1.1pt]{-}
{\bf Email ID} & \multicolumn{3}{l}{{\bf Top 3 User Reactions (\% of Users)}} \\
\hline
P-Greed & Delete (43\%) & {\bf Click} (32\%) &  NoAction (24\%)  \\
\hline
P-Curiosity &  {\bf Click} (44\%) &  NoAction (22\%) &  Delete (20\%)  \\
\hline
P-Urgency &  {\bf Click} (37\%) &  Delete (22\%) &  Visit (22\%)  \\
\hline
P-Fear &  Contact (48\%) &  {\bf Click} (20\%) &  Delete (18\%)  \\
\hline
Legit-1 &   Delete (38\%) &  NoAction (31\%) & {\bf Click} (26\%)  \\
\hline
Legit-2 &   Delete (45\%) &  {\bf Click} (26\%) &  NoAction (24\%)  \\
\hline
Legit-3 &   NoAction (31\%) &  {\bf Click} (25\%) &  Delete (23\%)  \\
\hline
Legit-4 &   Reply (46\%) &  {\bf Click} (31\%) &  NoAction (20\%)  \\
\tabucline[1.1pt]{-}
\end{tabu}
\end{center}
\vspace{-0.15in}
\end{table}

\begin{figure}[t]
\begin{center}
\includegraphics[width=0.4\textwidth]{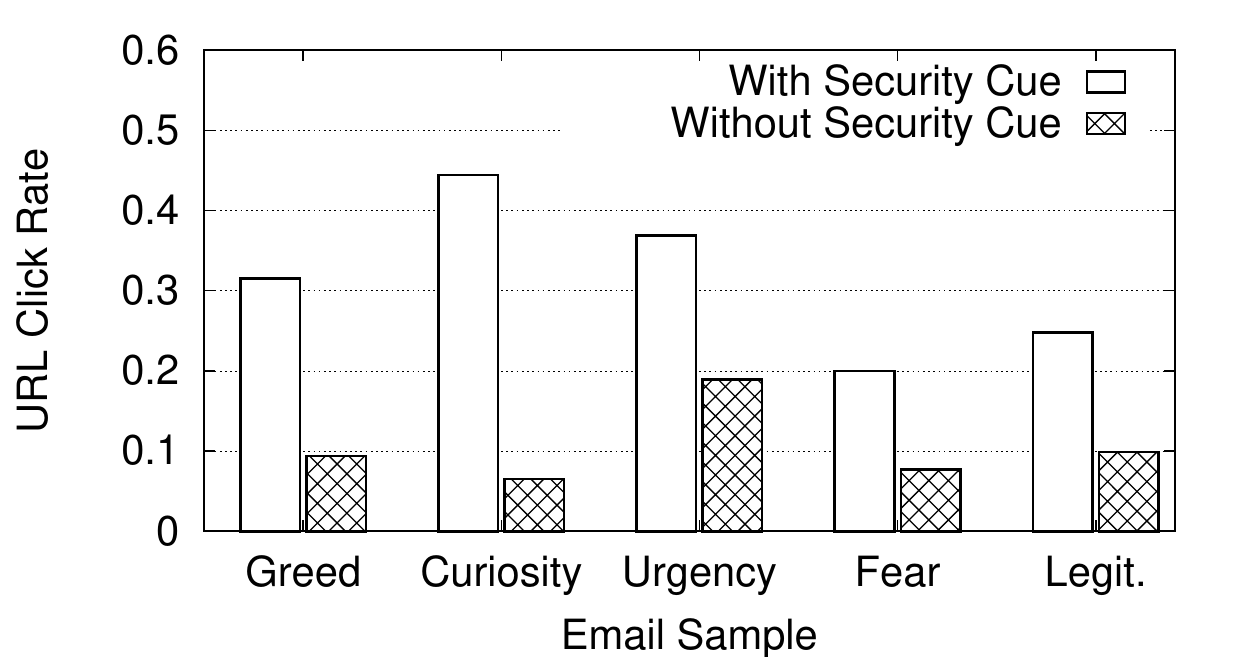}
\vspace{-0.15in}
\end{center}
\caption{Click-through rate of phishing URLs in study 1. ``URL click''
refers to users selecting ``Clicking on the URL'' option in the survey.}
\label{fig:click}
\vspace{-0.12in}
\end{figure}

\begin{table}[t]
\centering
\caption{Study 2 statistics.}
\label{tab:2surveyclick}
\vspace{-0.1in}
\begin{tabu}{l|l|l|l}
\tabucline[1.1pt]{-}
{\bf Phase} & {\bf Users} & {\bf Without Cue} & {\bf With Cue}  \\
\hline
\multirow{2}{*}{Phase1} & All Participants & 243 & 245 \\
&Not Blocking Pixel & 176 & 179 \\
\hline
\multirow{3}{*}{Phase2} & Opened Email & 94 & 86\\
& Clicked URL & 46 & 32 \\
\cline{2-4} 
& Click Rate & 48.9\% & 37.2\% \\
\tabucline[1.1pt]{-}
\end{tabu}
\vspace{-0.12in}
\end{table}

\begin{figure*}
\centering
    \subfigure[Role Playing]{
      \includegraphics[width=0.45\textwidth]{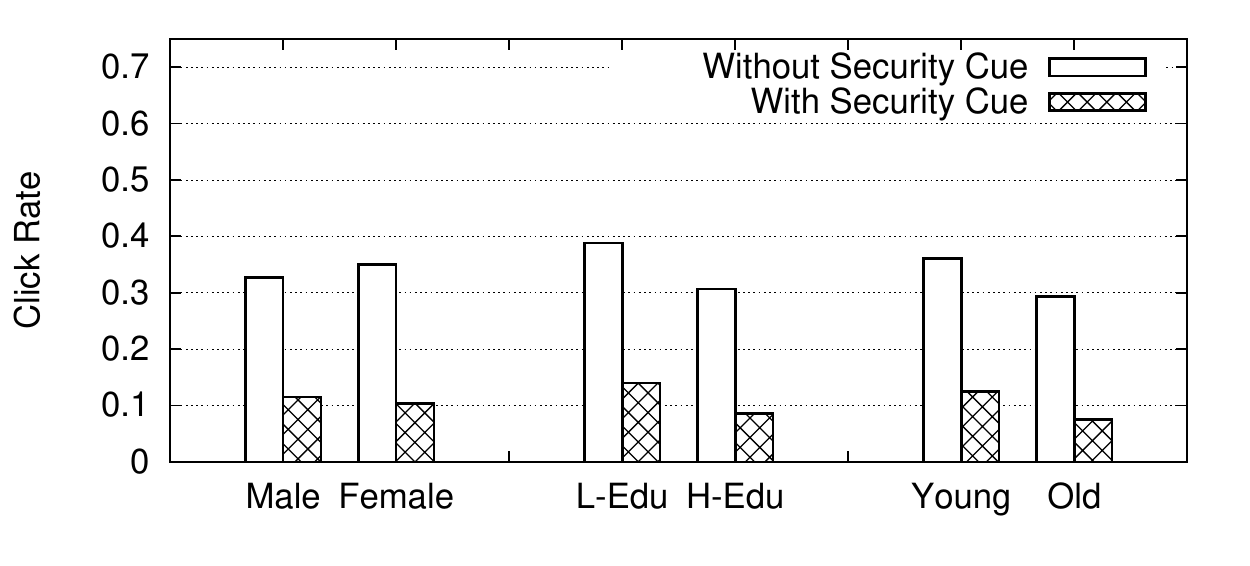}
       \vspace{-0.15in}
      \label{fig:demo1}
      \vspace{-0.05in}
    }
    \hfill
    \subfigure[Real Phishing]{
      \includegraphics[width=0.45\textwidth]{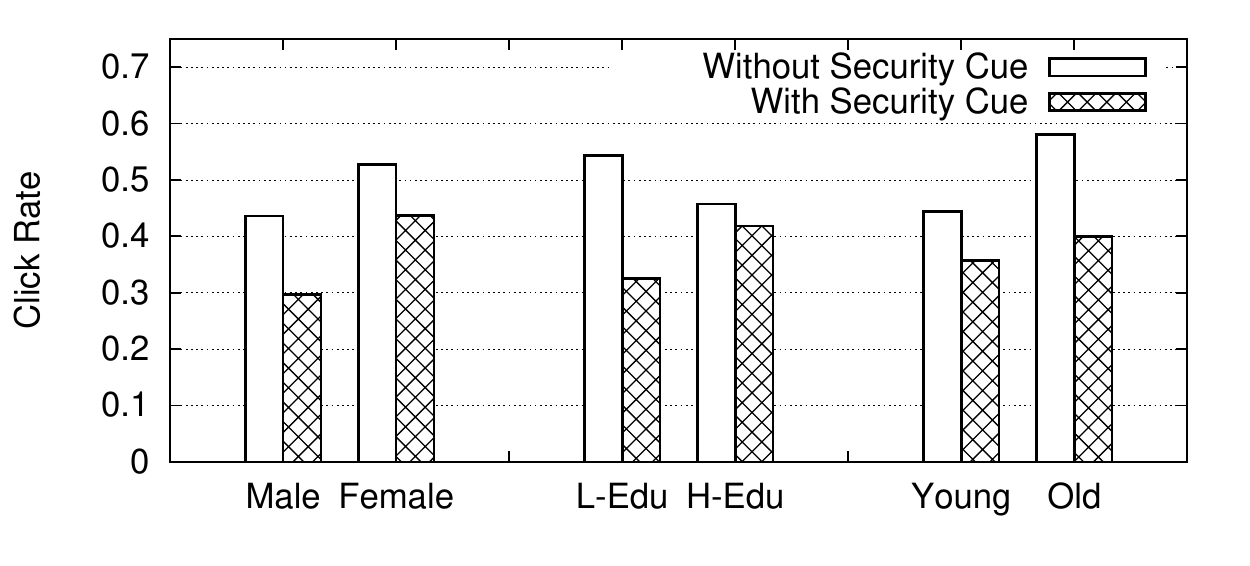}
     \vspace{-0.15in}
      \label{fig:demo2}
      \vspace{-0.05in}
    }
\vspace{-0.1in}
  \caption{The joint impact of demographic factors and security cues
    on click rates. For age, we divide users into Young (age$<$40) and
  Old (age$>=$40); For education level, we divide users into High-Edu
  (bachelor degree or higher) and Low-Edu (no bachelor degree). The
  thresholds are set so that the two compared groups have relatively
  even sizes.}
\vspace{-0.12in}
  \label{fig:demo}
\end{figure*}

\begin{figure}[t]
\begin{center}
\includegraphics[width=0.4\textwidth]{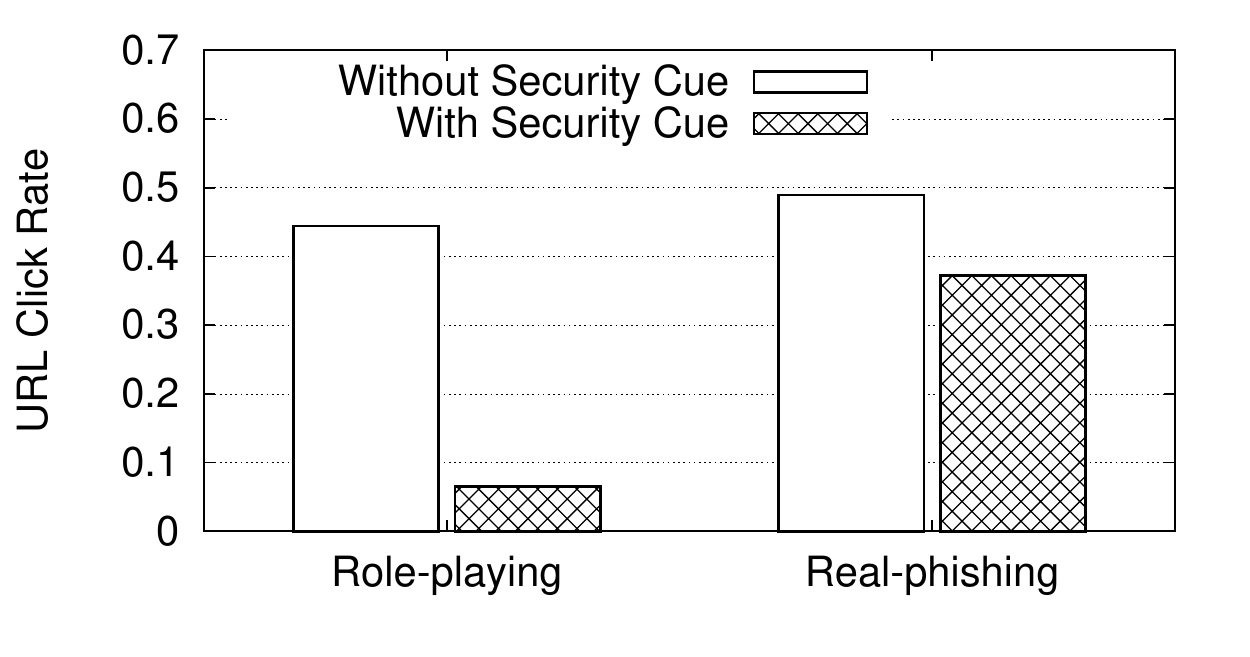}
\vspace{-0.17in}
\end{center}
\caption{Click-through rate of {\em P-Curiosity} in study 1 (role-playing) and study
  2 (real-phishing). }
\label{fig:phishme}
\vspace{-0.17in}
\end{figure}

\subsection{Results from Role-Playing Experiments}
To establish a baseline, we first analyze user behavior
under {\em no security cues}. Table~\ref{tab:react} shows the top 3 user reactions to
phishing and legitimate emails. We observe that user action is
highly dependent on the email content. For example, {\tt P-Curiosity}
and {\tt P-Urgency} are most effective in tricking users to click on the
URL in the email (the riskiest action). For {\tt P-Greed} and {\tt
  P-Fear}, ``clicking'' is the second most common action.
The result demonstrates the effectiveness of phishing emails.


Figure~\ref{fig:click} shows the ``click-through rate'' of the URLs
when security cues are presented. Note that for {\em study 1}, the ``clicking'' is
simulated, which refers to users selecting the ``click on the URL''
option in the survey. We observe that the security cue has a
clear impact on the user action. Comparing with the groups ``without
security cues'', the click-through rate of those ``with security cue''
drops by 12.3\%--37.9\%. The examine the statistical significance of
the observed differences, we run the Fisher's exact test\footnote{We
  also run the Chi-Square test and the results are
  consistent. In the rest of the paper, if not otherwise stated, we
  omit the Chi-Square result for brevity.} for the two
controlled groups (``with cue'' and ``without cue'') and the outcome (``clicked''
and ``not clicked''). Across all 4 email types, the differences are
significant ($p<0.05$). The results suggest that security cues have a significant impact to reduce the user
tendency of clicking on phishing URLs. In addition, Figure~\ref{fig:click} shows that false alerts
also reduce URL clicks --- false alerts do not
introduce direct harms to users. The negative impact might be
in the long run: if users are constantly exposed to false
alerts, users are more likely to ignore similar
alerts~\cite{DBLP:conf14d}. 

\subsection{Results from Real-world Phishing Tests}
For {\em study 2}, the click-through rate should
be computed based on users who {\em opened the email} instead of all
the users. This is because some participants did not check their inbox
during our monitoring period, and did not have the chance to see the security cue.
To evaluate the impact of the security cue, we first identify users
who opened the email based on the tracking
pixel. As shown in Table~\ref{tab:2surveyclick}, 176 and
179 users did not block tracking pixels in {\em phase1}. During {\em phase2},
94 and 86 of them have opened the email. 


The results indicate that security cues have a positive impact to
reduce phishing risks. When the security cue is presented,
the click rate is numerically lower (37.2\%) compared to that without
security cues (48.9\%). This result is consistent with {\em study 1}. In addition, we observe that the
difference between the two user groups in {\em study 2} is not as big
as that in {\em study 1} as shown in Figure~\ref{fig:phishme}. Fisher's exact test for {\em
phase 2} shows that $p=0.1329$, indicating the difference is
insignificant. Under realistic phishing experiments, the impact of the 
security cue is not as strong as that in the role-playing
setting. This may attribute to the fact that users are caught
off guard in {\em phase 2}. 

\subsection{Demographic Factors}
Finally, we cross-examine the results with respect to the demographic
factors in Figure~\ref{fig:demo}. For the role-playing statistics, we
aggregate the results for all 4 types of email content. To make sure each demographic group contains enough
users, we create binary groups for each factor. For ``education level'', we
  divide users into High-Edu (bachelor degree or higher) and Low-Edu
  (no bachelor degree). For ``age'', we divide users into Young (age$<$40) and
  Old (age$>=$40). The thresholds are chosen so that the two groups
  are of relatively even sizes. As shown in Figure~\ref{fig:demo}, the click
  rates are consistently lower when a security cue is presented for
  all the demographic groups. The differences are statistically significant for all the role-playing
  settings ($p<0.05$ based on Fisher's test). The differences for
  real-world phishing are insignificant (the smallest $p=0.06$ which
  is produced by the {\em low-edu} group). The result confirms the
  positive impact of security cue across different user demographics. 
  The statistically significant result from the role-playing
  experiment suggests that security cues have the potential to improve
  user efficacy in detecting phishing emails. The result from the
  real-world phishing tests indicates that the impact of security cues is not
  as strong when users are caught
  off-guard. Further research is needed to improve the design of security
  cues to maximize its positive impact.

%% file: discuss.tex
\section{Discussion}
\label{sec:discuss}

\para{Email Availability vs. Security.} Our study shows
many email providers choose to deliver a forged email to the inbox
even when the email fails the authentication. This is a difficult trade-off between
security and email availability. If an email provider blocks all the
unverified emails, users are likely to lose their emails ({\em
  e.g.}, from domains that did not publish an
SPF, DKIM or DMARC record). Losing legitimate emails is unacceptable for email
services which will easily drive users away. The current challenge is
the slow adoption rates of these anti-spoofing protocols among the
Internet and MX domains. 


\para{Countermeasures.} If the email providers decide
to deliver an unverified email to the inbox, we believe it is
necessary to place a security cue to warn users. Our user study shows
a positive impact of security cues in reducing risky user
actions. Another potential benefit of implementing security cue is that it can
act as a forcing function for sender domains to configure their
SPF/DKIM/DMARC correctly. 

There are other fixes that email providers (and protocol designers) may consider. 
First, email providers should consider adopting SPF, DKIM and
DMARC. Even though they cannot authenticate all the incoming emails,
they allow the email providers to make more informed decisions. The
current challenge is that these protocols have their own usability issues and limitations
(documented in~\cite{spf, dkim, dmarc}). Extensive work is needed to
make them easier to deploy and configure, to avoid major disruptions to
the existing email operation~\cite{yahoo14}. 


Second, email providers should make the security cues
{\em consistently} for different interfaces. Currently,
mobile users are exposed to a higher-level of risks due to the lack of
security cues. Another example
is that Google Inbox (web) users are less protected compared to Gmail users.
Third, misleading UI elements such as ``profile
photo'' and ``email history'' should be disabled for emails with
unverified sender addresses. 


\para{Questions Moving Forward.} Security cues help to reduce the phishing risk but
cannot eliminate the risk completely. Further research is needed to
understand how to design more effective cues to maximize its impact on
users. Another related question is how to maintain the long-term effectiveness of security
cues and overcome ``warning fatigue''~\cite{DBLP:conf14d}. 
For security-critical users ({\em e.g.}, journalists, military
personnel), security cues can only be a complementary solution. 
A better alternative is to use end-to-end encryption schemes such as
PGP~\cite{Garfinkel:1996:}. Further efforts are still needed to make
PGP widely accessible and usable for the broad population~\cite{Gaw:2006:SFP, google17}.

\para{Study Limitations.} Our study has a few limitations. First, our
measurement focuses on public email services. 
Future work will explore if the conclusion also applies to non-public email
services. 
Second, while we have taken significant efforts to
maintain the validity of the user study, there are still limits to
what we can control. For example, in study 1, we use a non-security
task to distract the participants. It is possible that some participants may still infer that security is the focus of the study.
In study 2, we only perform the user study on Yahoo Mail users with a
focus on a text-based cue. Our future work will look into expanding
the experiments to cover other email services and explore different security cue designs ({\em e.g.},
color, font, the wording of the message). 
Third, we use
``clicking on the phishing URL'' as a measure of risky actions, which
is still not the final step of a phishing attack. However, tricking users to give way their actual
passwords would have a major ethical implication, and we decided not to pursue
this step.

%% file: related.tex
\section{Related Work}
\label{sec:related}

\para{Email Confidentiality, Integrity and Authenticity.}
SMTP extensions such as SPF, DKIM, DMARC and STARTTLS are used to
provide security properties for email transport. Recently, researchers conducted
detailed measurements on the {\em server-side} usage of these
protocols~\cite{Durumeric:2015, Foster:2015, ndss:2016:}. Unlike prior
work, we focus on the {\em
  user-end} and demonstrate the gaps between server-side spoofing
detection and the user-end notifications. Our study is
complementary to existing work to depict a more complete picture.



\para{Email Phishing.} Prior works have developed phishing
detection methods based on features extracted from email content and
headers~\cite{DBLP:conf/ecrime, spear-phishing1, Fette:2007:L, Hong:2012, McGrath:2008:, Prakash:2010:}. 
Phishing detection is different from spam
filtering~\cite{Ramachandran:2007:} because phishing emails are not
necessarily sent in bulks~\cite{6289402} but can be highly
targeted~\cite{usenix17_spearphishing}. Other than spoofing, attackers may also apply
typosquatting or unicode characters~\cite{ndss15-typo} to make the
sender address {\em appear similar} (but not identical) to what they
want to impersonate. Such sender address is a strong indicator of
phishing which has been used to detect phishing
emails~\cite{Krammer:2006:,Kumaraguru:2007:}. 
Another line of research focuses on the {\em phishing website}, which
is usually the landing page of the URL in a phishing
email~\cite{Cui:2017:TPA:,Han:2016:PLM:,Vargas2016KnowingYE, 35580,
  Yue:Phinding,zhang2007cantina}. 

Human factors (demographics,
personality, cognitive biases, fatigue) would affect users
response to phishing~\cite{Oliveira:2017, 6957309, Jagatic:2007, jJPMB12, Sheng:2010:, Vishwanath:2011, Wang:2016, Wu:2006:, soups_bank, soups_children}. 
While most of these studies use the ``role-playing'' method, there are rare exceptions
~\cite{Jagatic:2007, Oliveira:2017} where the researchers conducted a real-world
phishing experiment. 
Our work is the first to examine the impact of security cues on
phishing emails, using both role-playing and realistic phishing
experiments. Our results show the different behavior of users in the role-playing and
real-world settings. An early work also demonstrates behavioral
differences in the role-playing experiments in online
authentication scenarios~\cite{ieeesp07}.



\para{Visual Security Cues. } Security cues are commonly used
in web or mobile browsers to warn users of insecure web
sessions~\cite{SOUPS16indicator:, SOUPS16:, Sunshine:2009:, luo2017hindsight}, 
phishing webpages~\cite{Dhamija, Egelman:2008, Wu:2006:, Zhang:2014:ESW:2556288.2557347}, and malware sites~\cite{Akhawe:2013:}. 
Existing work shows that users often ignore the security cues due to the lack
of understanding of the attack~\cite{Wu:2006:} or the frequent
exposure to false alarms~\cite{Krol:CRiSIS12}. Researchers have
explored various methods to make security UIs harder to ignore such as
using attractors~\cite{Bravo-Lillo:SP11, Bravo-Lillo:SOUPS14, Bravo-Lillo:SOUPS13}.
Our work is the first to measure the usage and effectiveness of security cues on forged
emails. We find that although security cues cannot completely
eliminate users' risky actions, they help to reduce such risk.








%% file: conclusions.tex
\section{Conclusion}
Through extensive measurements, controlled
user studies and real-world phishing tests, our work reveals a
concerning gap between the server-side email spoofing detection and
the actual protection on users. We demonstrate that most email
providers allow forged emails to get to user inbox, while lacking the
necessary warning mechanism to notify users (particularly on mobile
apps). For the few email services that implemented security cues, we
show that security cues have a positive impact on reducing risky user
actions under phishing attacks but cannot eliminate the risk. 
Moving forward, we believe more effective protections are needed to
defend users against spoofing and phishing attacks. 

%% file: appendix.tex

\section*{Appendix A -- SPF/DMARC Adoption}
We examine the adoption rate of SPF and DMARC by crawling the DNS record for Alexa top 1
million hosts. Table~\ref{tab:jan} shows statistics for January 2017.

\begin{table}[th]
\begin{center}
\caption{SPF/DMARC statistics of Alexa 1 million domains.}
\label{tab:jan}
\begin{tabu} to 0.47\textwidth{X[l]X[r]X[r]}
\tabucline[1.1pt]{-}
 Status  & All Domains \# (\%)  & MX Domains \# (\%)  \\
\hline
Total domains & 1,000,000 (100\%) & 818,682 (100\%) \\
\hline
w/ SPF & 477,484 (47.7\%) & 461,402 (56.4\%) \\
w/ valid SPF & 433,640 (43.4\%) & 418,284 (51.1\%) \\
Policy: soft fail & 268,886 (26.9\%) & 264,185 (32.3\%) \\
Policy: hard fail  & 110,225 (11.0\%) & 100,665 (12.3\%) \\
Policy: neutral & 53,218 (5.3\%) & 52,150 (6.4\%) \\
Policy: pass & 1,311 (0.1\%) & 1,284 (0.2\%) \\
\hline
w/ DMARC & 30,996 (3.1\%) & 29,681 (3.6\%) \\
w/ valid DMARC &  30,594 (3.1\%) &  29,310 (3.6\%) \\
Policy: none &  23,336 (2.3\%) &  22,728 (2.8\%) \\
Policy: reject & 4,199 (0.4\%) &  3,581 (0.4\%)\\
Policy: quarantine & 3,059 (0.3\%) & 3,001 (0.4\%) \\
\tabucline[1.1pt]{-}
\end{tabu}
\end{center}
\vspace{-0.25in}
\end{table}

\begin{table}[h]
\centering
\caption{Shadow Experiment Domain List.}
\label{appendix:shadow}
\vspace{-0.05in}
\begin{tabu}{l}
\tabucline[1.1pt]{-}
No SPF/DKIM/DMARC (20) \\
\hline
careerride.com, gmw.cn, sunrise.ch, office.com, thepiratebay.org, \\
onlinesbi.com, thewhizmarketing.com, askcom.me, putlockers.ch, \\
spotscenered.info, twimg.com, thewhizproducts.com,  reddituploads.com, \\
codeonclick.com, ltn.com.tw, 4dsply.com, mysagagame.com, \\ 
ablogica.com, sh.st, mixplugin.com \\\hline \hline
SPF/DKIM;DMARC=none (20) \\
\hline
tumblr.com, wikipedia.org, ebay.com, msn.com, imgur.com, apple.com, \\
github.com, microsoftonline.com, coccoc.com, adf.ly, alibaba.com, \\
bbc.co.uk, rakuten.co.jp, rakuten.co.jp, vimeo.com, nytimes.com, \\
espn.com,  salesforce.com, godaddy.com, indiatimes.com, weather.com\\ \hline \hline

SPF/DKIM;DMARC=reject (20) \\
\hline
youtube.com, google.com, pinterest.com, vk.com, twitter.com, \\
instagram.com, reddit.com, linkedin.com, blogspot.com, microsoft.com, \\
ok.ru,  stackoverflow.com,  netflix.com, paypal.com, twitch.tv, adobe.com, \\
dropbox.com, whatsapp.com, booking.com, stackexchange.com\\
\tabucline[1.1pt]{-}
\end{tabu}
\vspace{-0.1in}
\end{table}

\begin{figure}[h]
\centering    
    \subfigure[Google Inbox profile photo (same-domain spoofing)]{
      \includegraphics[width=0.345\textwidth]{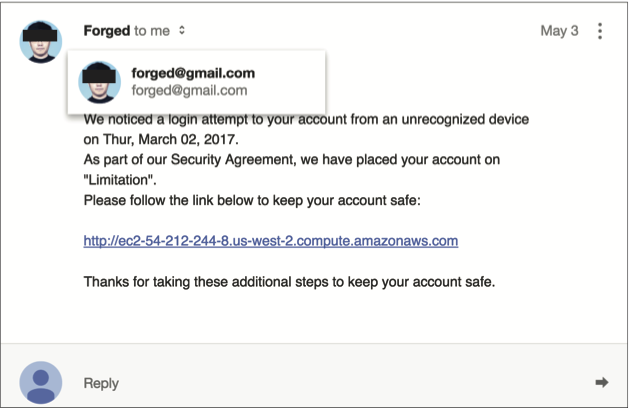}
       \vspace{-0.05in}
      \label{fig:ui1}
      \vspace{-0.1in}
    }
    \vfill
    \subfigure[Seznam profile photo (same-domain spoofing) ]{
      \includegraphics[width=0.345\textwidth]{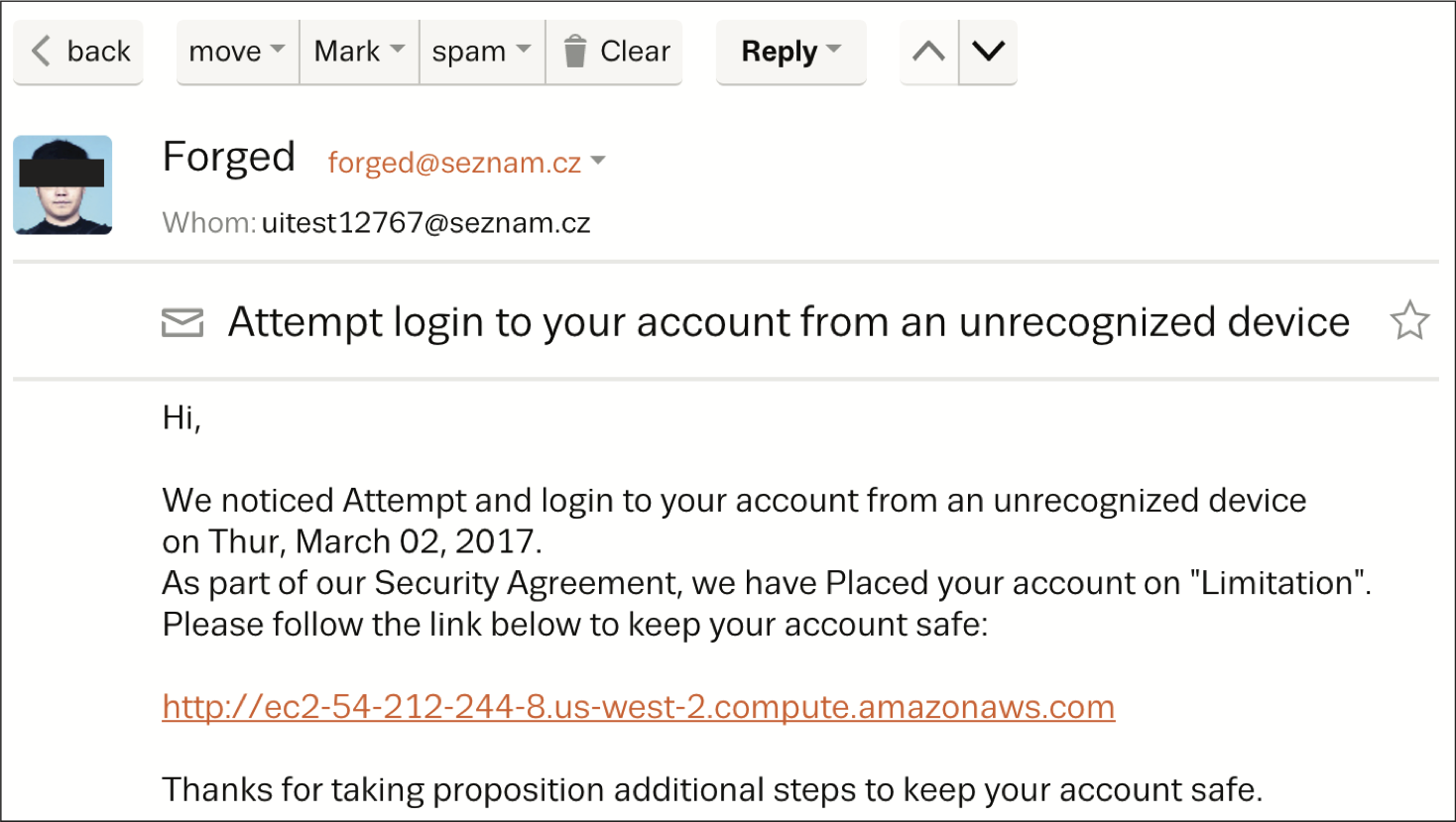}
     \vspace{-0.05in}
      \label{fig:ui2}
      \vspace{-0.1in}
    }
    \vfill
    \subfigure[Zoho profile photo and email history (spoofing a contact)]{
      \includegraphics[width=0.345\textwidth]{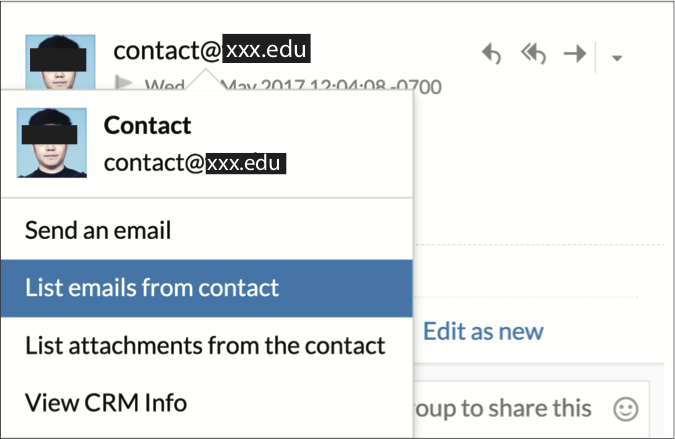}
     \vspace{-0.05in} 
      \label{fig:ui3}
      \vspace{-0.1in}
    }
\vspace{-0.05in}
  \caption{Examples of misleading UIs (profile photo, email history,
    namecard).
  }
\vspace{-0.25in}
  \label{fig:ui}
\end{figure}

\section*{Appendix B -- Shadow Experiment}
Table~\ref{appendix:shadow} lists the 60 domains used by the end-to-end
spoofing experiment (shadow experiment). The domains per category are
selected from Alexa top 2000 domains.

\section*{Appendix C -- Misleading User Interface}
Figure~\ref{fig:ui} shows three examples of misleading UI elements. 
Figure~\ref{fig:ui1} and \ref{fig:ui2} show that when an attacker spoofs a user from the same email provider as the
receiver, the email provider will automatically load the profile {\em
  photo} of the spoofed sender from its internal database. In both Google
Inbox and Seznam, the forged emails look like that they were sent by the
user ``Forged'', and the photo icon gives the forged email a more
authentic look.

Figure~\ref{fig:ui3} demonstrates the misleading UIs when the attacker
spoofs an existing contact of the receiver. Again, despite the sender
address ({\tt contact@xxx.edu}) is spoofed, Zoho
still loads the contact's photo from its internal database. In addition, users can check the recent email conversations
with this contact by clicking on the highlighted link. These elements
make the forged email look authentic.  The profile photos used the
examples are not from the authors, but we received the permission from
the photo owner. Eyes are obscured for the paper submission.


\section*{Appendix D -- Phishing Email Screenshots}
\begin{figure}[ht]
\centering    
    \subfigure[Without Security Cue]{
      \includegraphics[width=0.45\textwidth]{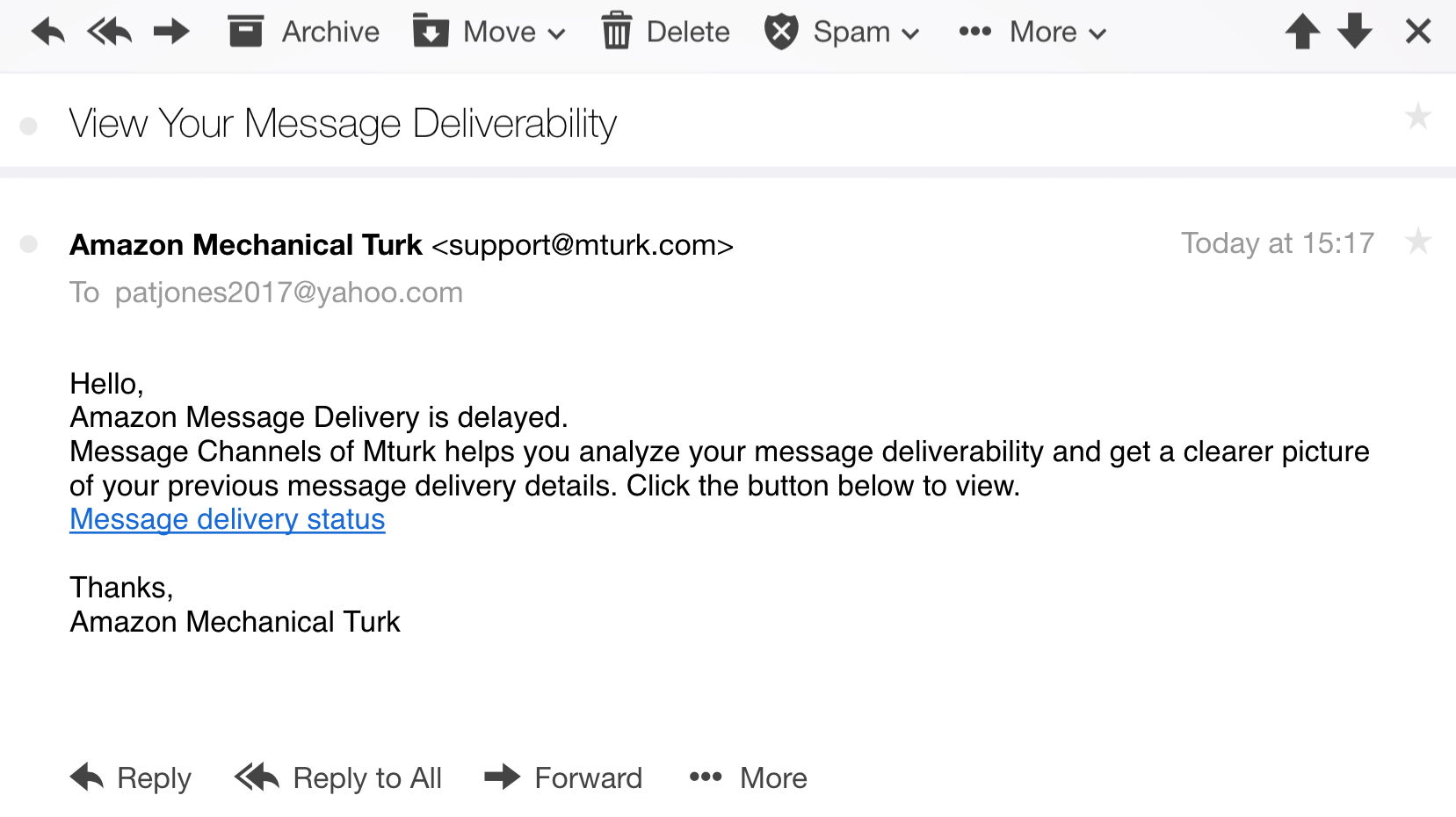}
       \vspace{-0.05in}
      \label{fig:phish1}
      \vspace{-0.05in}
    }
    \vfill
    \subfigure[With Security Cue]{
      \includegraphics[width=0.45\textwidth]{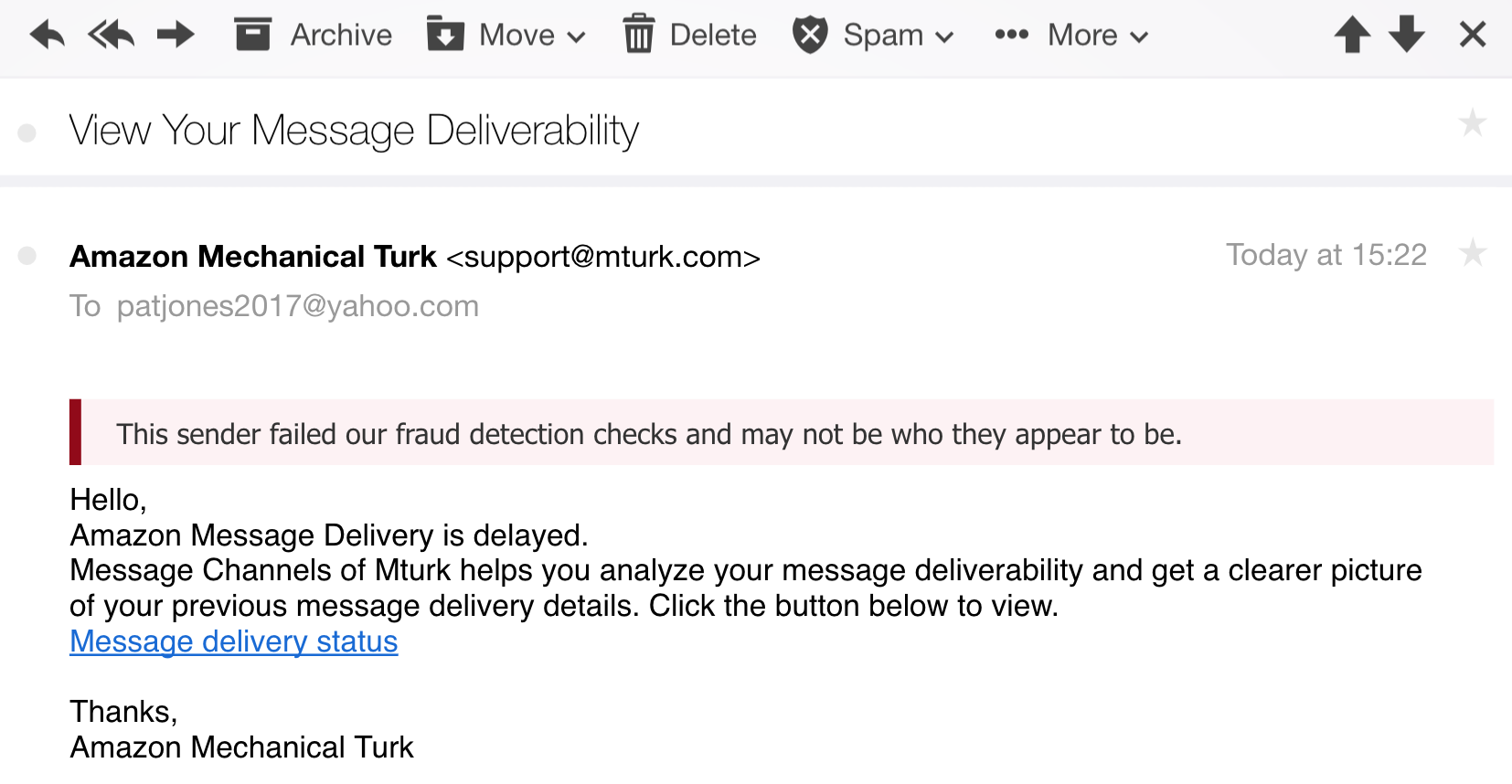}
     \vspace{-0.05in} 
      \label{fig:phish3}
      \vspace{-0.05in}
    }
\vspace{-0.05in}
  \caption{The phishing email screenshot. 
  }
\vspace{-0.05in}
  \label{fig:phish}
\end{figure}

Figure~\ref{fig:phish} 
shows the example phishing email used in
our user studies (for both study 1 and study 2). The email content is {\tt
  P-Curiosity}. These are screenshots taken from the inbox of Yahoo
Mail, reflecting what the participants saw in the phishing
experiments.

